\documentclass[aps,pra,reprint,superscriptaddress,nofootinbib]{revtex4-1}
\usepackage{amsmath,amssymb,bm}
\usepackage{graphicx}
\usepackage{hyperref}
\usepackage{xcolor}
\begin{document}

%\title{Multimode Liouvillian Encoding of Noise-Induced Coherence Beyond the Liouvillian Gap}
%\title{Quantum Fisher Information of Noise-Induced Coherence Beyond the Liouvillian Gap through Multimode Dynamics}
%\title{Quantum Fisher Information Matrix of Coherence Beyond Liouvillian Gap }
\title{ Coherence Estimation Beyond the Liouvillian Gap in a Finite Nonequilibrium  System }
%\title{Quantum Fisher Information of Noise-Induced Coherence Beyond the Liouvillian Gap}
\author{Sonali Brahma}
\author{Trishna Kalita}
\author{Himangshu Prabal Goswami}
\email{hpg@gauhati.ac.in}
\affiliation{QuAinT Research Group, Department of Chemistry, Gauhati University, Jalukbari, Guwahati-781014, Assam, India}
\date{\today} 

\date{\today}

\begin{abstract}
We investigate the estimation of bath-induced coherence in a finite quantum system interacting with thermal reservoirs.  Enhancement of coherence estimation is transient and the estimation precision totally disappears at the steady state despite the system retaining finite coherence. By analyzing the full Liouvillian eigenspectrum, we demonstrate that the optimal sensing window emerges from the competition between identifiable contributory modes' temporal relaxation and statistical importance. Neither is the linear inverse scaling of Liouvillian gap with transient optimal time a signature of unimodal contribution to optimal sensing, nor is the existence of multimodal dynamics a signature of nonlinear scaling.  The inverse Liouvillian gap does not obey any general scaling with the optimal sensing time of coherence and we prove our numerical results analytically using a general Markovian framework. We further show that coupling the finite system to a quantum cavity and maintaining a  thermal bias, transforms the transient metrological optimization into a sustained steady-state resource.  
\end{abstract}


\maketitle

\section{Introduction}
Quantum parameter estimation aims to determine physical parameters with the highest achievable precision. The ultimate precision is bounded by the quantum Cram\'er-Rao bound, where the quantum Fisher information (QFI) quantifies the statistical distinguishability of neighboring quantum states \cite{Braunstein1994,Paris2009,Giovannetti2004,Giovannetti2011,Demkowicz2015}. In realistic settings,  parameter estimation must be formulated within an open-system framework, where unavoidable environmental interactions modify the dynamics \cite{Escher2011,Alipour2014}. Although decoherence generally degrades precision, engineered dissipation can itself become a metrological resource through reservoir engineering, dissipative state preparation, autonomous error correction, and dissipative sensing \cite{Verstraete2009,Diehl2008,Poyatos1996}. Consequently,  QFI analysis has been extended to multipartite entanglement, quantum criticality, and quantum phase transitions \cite{Pezze2018,Gu2010}.

Traditionally, open-system dynamics are usually explored through the Liouvillian superoperator $\mathcal{L}$, which incorporates both coherent evolution and dissipation and determines the hierarchy of relaxation processes toward the unique steady state \cite{Minganti2018,Kessler2012}.  The Liouvillian gap ($\Delta$), defined as the smallest nonzero real part of the Liouvillian spectrum, sets the slowest relaxation timescale and is therefore closely related to information accumulation during sensing. Likewise, the parameter susceptibility depends on the Liouvillian pseudoinverse, whose norm has is known to scale linearly with the inverse gap. Consequently, a closing Liouvillian gap can amplify the steady-state response to perturbations and enhance the QFI, as demonstrated near the dissipative superradiant transition of the open Dicke model \cite{Wang2014} and in finite-size open Dicke systems \cite{Yu2022}. Similar inverse-gap scaling accurately predicts long-time relaxation in boundary-dissipated Anderson-localized chains \cite{Zhou2022}. Since gap closing is a hallmark of dissipative phase transitions and nonequilibrium criticality, often accompanied by enhanced fluctuations and metrological sensitivity \cite{Vicentini2018,Carollo2020}, it is frequently assumed to determine the timescale for information accumulation and thus the optimal interrogation time as the transient buildup of Fisher information \cite{Thakuria2026,Saleem2023} is expected to be governed by the slowest relaxation mode of the Liouvillian.

An expectation of simple linear relationships between metrological quantities  with the Liouvillian's quantities  is not generally valid because the QFI depends on the full Liouvillian spectral decomposition, including eigenmode amplitudes, overlaps with parameter perturbations, and non-normal effects. Relaxation can substantially exceed $1/\Delta$ when transient dynamics are governed by multiple decay modes rather than the slowest eigenvalue alone. In boundary-driven spin chains, relaxation is controlled by superexponentially large expansion coefficients of higher Liouvillian modes rather than the gap \cite{Mori2020}. The Liouvillian skin effect produces boundary-localized eigenmodes whose relaxation times can greatly exceed $1/\Delta$ despite a finite gap \cite{Haga2021}. Likewise, the symmetrized Liouvillian gap has been shown to characterize only the asymptotic exponential decay, not generic transient relaxation \cite{Mori2023}. Therefore, any inverse-gap scaling of the optimal sensing time should be regarded as an asymptotic or regime-dependent result rather than a universal consequence of pseudoinverse scaling. QFI enhancement additionally depends on the parameter dependence of the Liouvillian, the overlap of perturbations with slow decay modes, and the non-Hermitian spectral structure of the Liouvillian \cite{Peng2024,Midha2025}. Thus, although the gap remains an important dynamical indicator, optimizing estimation precision requires the full Liouvillian spectrum.

In a standard open quantum system, apart from commonly measured system parameters like coupling or energy or flux ,  appearance of coherence is also expected and precise measurement of coherence has always been a  challenge. In this work, we therefore focus on how precisely coherences in open quantum systems can be estimated by  computing the time-resolved QFI matrix. Specifically, we estimate noise-induced or bath induced coherence (BIC) \cite{Scully2003,Dorfman2013,Uzdin2015,Scully2011,Dorfman2018,Sarmah2024,Camati2019,Holubec2018}, a nonequilibrium coherence generated by quantum interference between dissipative pathways that produces steady-state coherence without coherent driving \cite{Scully2003,Dorfman2013}. The BIC parameters enter directly into the Liouvillian and their unknown metrological aspects can be explored. Our work therefore bridges quantum estimation theory with models traditionally employed to demonstrate thermodynamic advantages of coherence \cite{Scully2011,Dorfman2018}. We explicitly show, within a concrete model, when and which modes of the complete Liouvillian spectrum beyond the gap govern transient coherence estimation. We also analytically show that, for a general Markovian system, no simple scaling relationship exists between optimal sensing time and Liouvillian gaps.

The paper is organized as follows. First, we discuss the finite quantum system, a four level $V-$type system with two degenerate states and the emergence of bath induced coherence withing a master equation framework. The complete analytical Liouvillian with coupled population and coherences is also discussed. In the next section, we present our results and discussion of the transient optimization of the two coherence's QFI along with the correlations during simultaneous estimation. We perform a complete spectral decomposition of the density matrix in terms of the Liouvillian's eigenspectrum and establish the role of multimodal relaxation and statistical weights of the modal expansion amplitudes. Here we also show that a linear inverse scaling of Liouvillian gap and transient optimal time is not a signature of unimodal contribution to optimal sensing time. We also show that multimodal contribution can also lead to linear scaling between the two. In the subsequent section, we show how coupling the degenerate $V$-type system to a quantum cavity shifts the transient optimization to the steadystate. In the next section, we provide an analytical reasoning for a general Markovian system which show why linearity or nonlinearity between inverse Liouvillian gap and interrogation time can both be observed in the QFI, after which we conclude. 
\section{Nonequilibrium Model Dynamics}
 
\begin{figure}[ht]
\centering
\includegraphics[width=\columnwidth]{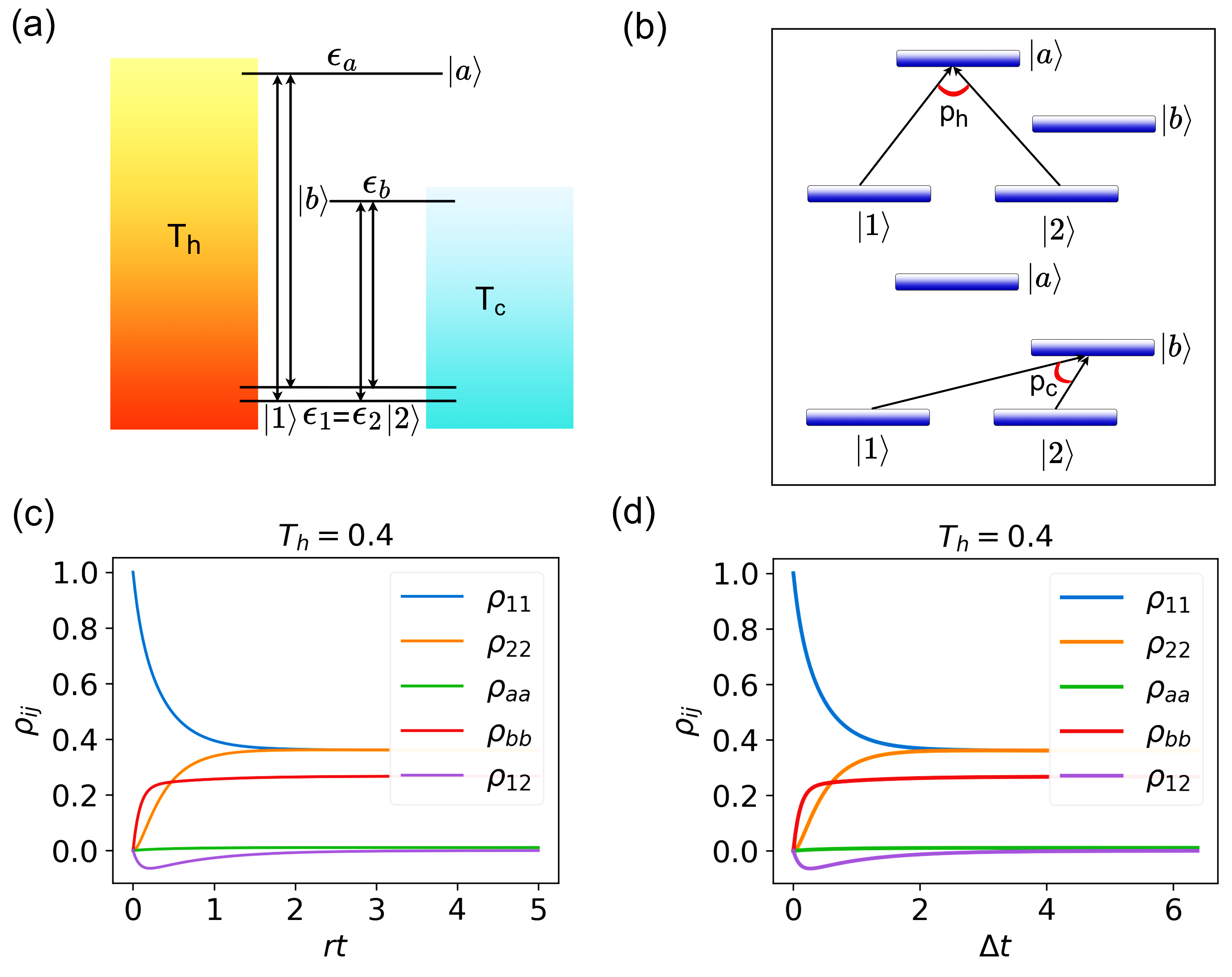}
\caption{(a) Schematic representation of the nonequilibrium four-level open quantum system. The degenerate ground states $|1\rangle$ and $|2\rangle$ are coupled to the excited states $|a\rangle$ and $|b\rangle$ through hot and cold thermal reservoirs at temperatures $T_h$ and $T_c$. System parameters are $\epsilon_{1}$ = $\epsilon_{2}$ = 0.1, $\epsilon_{a}$ = 1.5, $\epsilon_{b}$ = 0.4, $T_h$ = [0.4, 3], $T_c$ = 1, $T_l$ = 0.2, $k_B$ = 1, $r$ = 1. (b) Schematic representation of the origin of the hot-bath coherence parameter $p_h$, and cold bath coherence parameter $p_c$. (c) Time evolution of the states  $\rho_{11}$, $\rho_{22}$, $\rho_{aa}$, $\rho_{bb}$, and $\Re\rho_{12}$ in the regime. (d) Evolution of states  as a function of rescaled time $\Delta t$. In panel (c,d) parameters $p_h$ = 0.62, $p_c$ = 0.7. Natural units are used throughout, $k_B\to 1,\hbar\to 1$. } 
\label{model_population}
\end{figure}

 The nonequilibrium system that we consider is four-level quantum system consisting of two degenerate ground states, $|1\rangle$ and $|2\rangle$, coupled asymmetrically to two excited states, $|a\rangle$ and $|b\rangle$, through two thermal reservoirs as illustrated in Fig.~(\ref{model_population}a).  The hot and cold reservoirs generate incoherent excitation and relaxation processes giving rise to nonequilibrium steady-state coherence and rich transient dynamics. Such a model has been extensively studied as examples of artificial bioinspired quantum photocells \cite{Creatore2013}, heat engines \cite{Sarmah2024}, refrigerators \cite{Holubec2018,Holubec2019,Uzdin2015}, even served as a prototype for photosytnthetic reaction center \cite{Dorfman2013} and proven to lead to exotic physical insights.

The total Hamiltonian is written as
\begin{equation}
H_T
=
H_s
+
H_b
+
H_{sb}
\end{equation}
where $H_s$ and $H_b$ denote the system and reservoir Hamiltonians, respectively, while $H_{sb}$ and $H_{s\ell}$ describe the system-bath interactions. The system Hamiltonian is
\begin{equation}
H_s
=
\sum_{m\in\{1,2,a,b\}}
\epsilon_m
c_m^\dagger c_m,
\end{equation}
where $\epsilon_m$ denotes the energy of the corresponding level,
while each thermal reservoir is modeled as a continuum of harmonic modes,
\begin{equation}
H_b
=
\sum_{b\in\{c,h\}}
\sum_k
E_{bk}
a_{bk}^\dagger
a_{bk}.
\end{equation}
The interaction between the system and the thermal reservoirs is
\begin{equation}
H_{sb}
=
\sum_{b\in\{c,h\}}
\sum_k
\sum_{i=1,2}
\sum_{\alpha=a,b}
r_{ibk}
\left(
a_{bk}c_\alpha^\dagger
+
a_{bk}^\dagger c_\alpha
\right),
\end{equation}
where $r_{ibk}$ denotes the system-bath coupling strength.
Under the standard Born--Markov approximation, tracing out the reservoir degrees of freedom leads to the reduced master equation
$
\dot{\rho}
=
\mathcal{L}\rho,
$
where $\mathcal{L}$ is the Liouvillian superoperator governing the dissipative dynamics. When the couplings to the baths are assumed to be equal, the Liouvillian takes the matrix form 
\begin{equation}
\label{eq:Lmatrix}
\mathcal L =\begin{pmatrix}
- r n & 0 & r\tilde n_h & r\tilde n_c & - r y\\
0 & - r n & r\tilde n_h & r\tilde n_c & - r y\\
r n_h & r n_h & -2r\tilde n_h & 0 & 2r p_h n_h\\
r n_c & r n_c & 0 & -2r\tilde n_c & 2r p_c n_c\\
-\frac{ry}{2} & -\frac{ry}{2} & rp_h\tilde n_h & rp_c\tilde n_c & -rn
\end{pmatrix},
\end{equation}
where
$
n_i=
1/\{\exp[(E_b-E_1)/(k_BT_i)]-1\},
$
$\tilde n_i=1+n_i$, $n=n_c+n_h$, and $y=n_cp_c+n_hp_h$ in the reduced picture $\rho =\{\rho_{11}$, $\rho_{22}$, $\rho_{aa}$, $\rho_{bb}$, $\Re(\rho_{12})\}$.  Note that,  $p_h$ and $p_c$ are  two dimensionless parameters characterizing the strength of the bath-induced (noise-induced) coherence generated by the hot and cold reservoirs. As shown in Fig.~(\ref{model_population}a), the hot (cold) reservoir simultaneously couples the degenerate ground states $|1\rangle$ and $|2\rangle$ to the common excited state $|a\rangle$ ($|b\rangle$), allowing interference between the corresponding transition pathways as shown in Fig.~(\ref{model_population}b)  and thereby generating the ground-state coherence $\rho_{12}$. This interference is incorporated through the mixed system-reservoir coupling terms during the perturbative treatment of constructing the superoperator through pseudotransition rates associated with the two pathways \cite{Goswami2013,Sarmah2024,Dorfman2013,Scully2011}. The two quantities  are hence parametrized versions of the degree of coehernce and are related to the angle between the corresponding transition dipole moments associated with each coupled transition, Fig. (1b). $p_\nu=0$ corresponds to the absence noise-induced coherence, while $p_\nu=1$ corresponds to strongest generation of the coherence $\rho_{12}$. In this work $p_h$ and $p_c$ are treated as the  parameters to be estimated. The exact derivation of such Liouvillians in perturbative formalism has been reported in multiple works \cite{Dorfman2013,Sarmah2024, sarmah2024efficiency,Holubec2018}.

We now investigate the metrological properties of the nonequilibrium four-level system. The quantities of interest are the coherence parameters $p_c$ and $p_h$, which quantify the interference induced by the cold and hot reservoirs, respectively. Since both parameters enter explicitly through the Liouvillian, the transient density matrix inherits their dependence through the dissipative dynamics. The density matrix at every instant is obtained from the solution of the master equation which are shown in Fig.~(\ref{model_population}c,d). Note that the coherence $\Re \rho_{12}$ survives.

\section{Results and Discussion}

For a general Markovian master equation the Liouville space, 
the evolution of the density operator is governed by
\begin{equation}
\frac{d\rho(t)}{dt}
=
\mathcal{L}(\boldsymbol{\theta})\rho(t),
\label{eq:lindblad}
\end{equation}
and 
$\boldsymbol{\theta}=(\theta_1,\theta_2,\ldots)$ are parameters that are to be estimated. 
The symmetric logarithmic derivative (SLD, $L_\mu$) corresponding to the parameter
$\theta_\mu$ is defined through the Sylvester equation
\begin{equation}
\partial_\mu\rho
=
\frac12
\left(
L_\mu\rho
+
\rho L_\mu
\right),
\label{eq:sld}
\end{equation}
where $
\partial_\mu
\equiv
\frac{\partial}{\partial\theta_\mu}.$ The Quantum Fisher Information Matrix (QFIM) element is then given by
\begin{equation}
F_{\mu\nu}
=
\frac12
\operatorname{Tr}
\left[
\rho
\left(
L_\mu L_\nu
+
L_\nu L_\mu
\right)
\right].
\label{eq:qfim}
\end{equation}
The density operator in the biorthogonal eigenbasis of the Liouvillian is,
\begin{equation}
\rho(t)
=
\rho_{\mathrm{ss}}
+
\sum_{n>0}
c_n
e^{\lambda_nt}
r_n,
\label{eq:rho_spec}
\end{equation}
where $r_n$ and $l_n$ denote the right and left eigenoperators satisfying
$
\mathcal L r_n=\lambda_nr_n,
$
$
l_n^\dagger\mathcal L=\lambda_nl_n^\dagger,
$
and
$
\mathrm{Tr}(l_m^\dagger r_n)=\delta_{mn}.
$
The modal expansion coefficients, a measure of statistical importance or weights,
\begin{equation}
c_n
=
\frac{\mathrm{Tr}\!\left(l_n^\dagger\rho(0)\right)}
{\mathrm{Tr}\!\left(l_n^\dagger r_n\right)}
\label{eq:cn}
\end{equation}
measure the contribution of each relaxation mode to the dynamics. 
The modal expansion coefficients
are completely determined by the overlap of the initial state with the left and right Liouvillian eigenoperators and are not independent fitting parameters.
We consider the simultaneous estimation of the two coherence parameters $p_c$ and $p_h$, i.e the estimation parameter space is $\theta\in\{p_c,p_h\}$.
In this case, in the FQIM, $F_{\mu\nu}$,  the diagonal elements,
$F_{p_cp_c}$ and $F_{p_hp_h}$,
coincide with the single-parameter QFIs for $p_c$, $F_Q(p_c)$ and $p_h, F_Q(p_h)$, respectively. The off-diagonal elements of the FQIM, $F_{ch}$ and $F_{hc}$ quantify the statistical correlation between the two estimations. 

\begin{figure}[ht]
\centering
\includegraphics[width=\columnwidth]{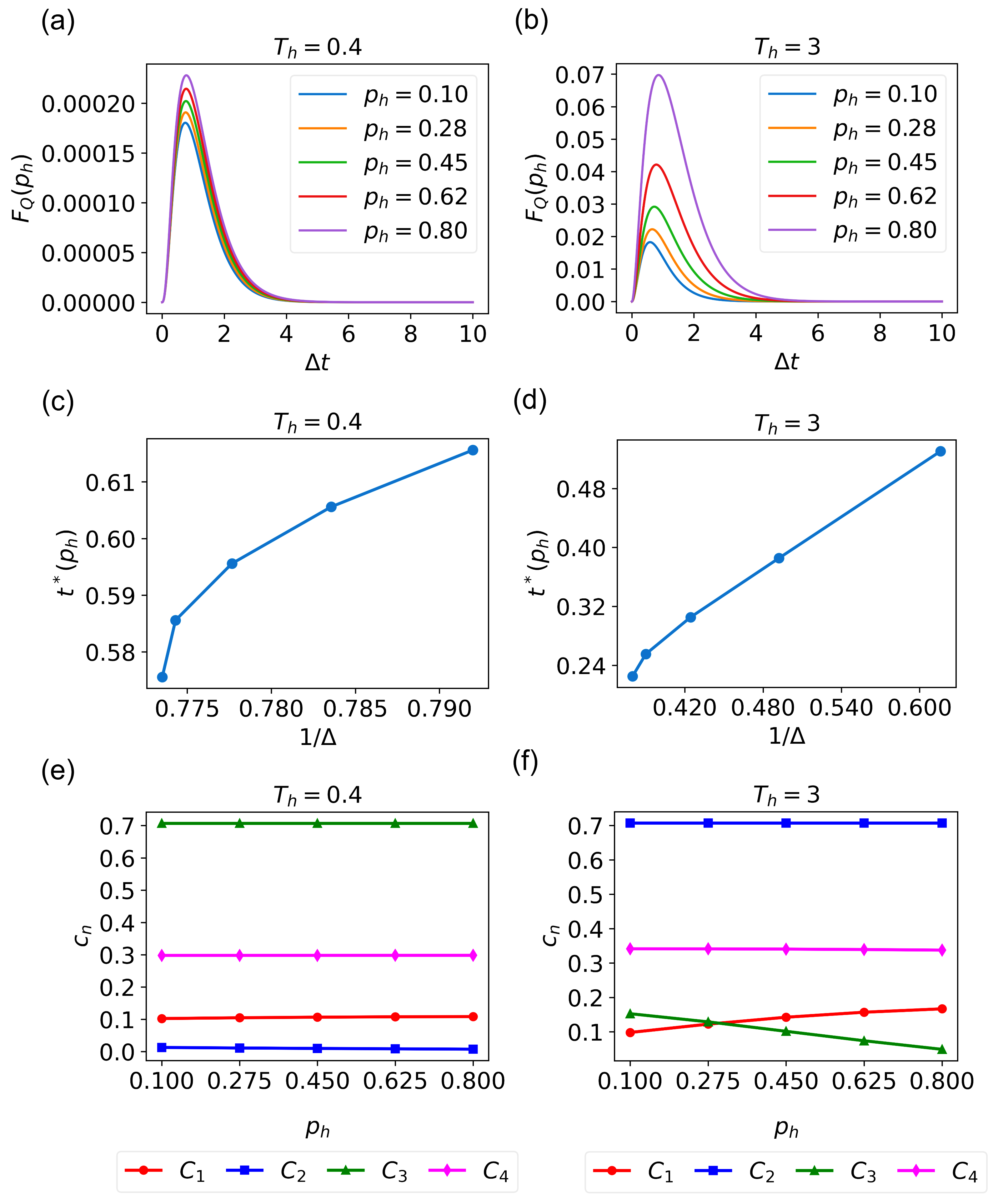}
\caption{(a,b) $F_Q(p_h)$ as a function of rescaled time $\Delta t$ with fixed $p_c$ = 0.7. (c,d) Optimal estimation time $t^*(p_h)$ as a function of the inverse of Liouvillian gap $\Delta$. In (c), one mode contributes and yet there is a nonlinearity. (e,f) $c_n$ as a function of $p_h$.}
\label{F_ph_C_n}
\end{figure}

In Fig.~(\ref{F_ph_C_n}a,b),  we numerically evaluate $F_{Q}(p_h)$ for several $p_h$ values as a function of $\Delta t$ for two different hot bath temperatures ($T_h = 0.4$ and 3) respectively. In each case $F_Q(p_h)$ grows from zero to a maximum and then decays to zero. Therefore,  the noise-induced coherence due to the hot bath, $p_h$ is best estimated at the transient state and not at the steady-state. Moreover, the zero value of $F_Q(p_h)$ at steady state implies that the coherence cannot be estimated with the greatest precision once the system reaches the steady-state despite being finite at the steady-state (the coherence term is finite in Fig. (1c,d)). The magnitude of $F_Q(p_h)$ dramatically increases when the bath temperature is increased , indicating greater precision of estimating $p_h$ at higher temperatures. Also note that, based on the magnitude of the QFI at different values of the coherence, higher the coherence, better it can be estimated.  We identify the peak time $t^*$ by finding $\partial_tF_Q=0$. 

Fig.~(\ref{F_ph_C_n}c) shows how $t^*$ varies the inverse Liouvillian gap $1/\Delta$  evaluated at different $p_h$ values at lower temperature. We observe a nonlinear relation.  At higher hot bath temperature, Fig.~(\ref{F_ph_C_n}d), the graph appears linear with a slight nonlinearity at lower $1/\Delta$ values. Typically, when the Liouvillian gap sets the dominant metrological clock  one  expects $t^*\propto 1/\Delta$, and if there is multimodal contribution or other modes at play, the $t^*$ should be nonlinear with $1/\Delta$.  However, we shall demonstrate that such appearance of linearity or nonlinearity should not be associated with an inference that a single Liouvillian gap or multimodes primarily contribute to QFI in the transient state respectively.  

\begin{figure}
\centering
    \includegraphics[width=1\linewidth]{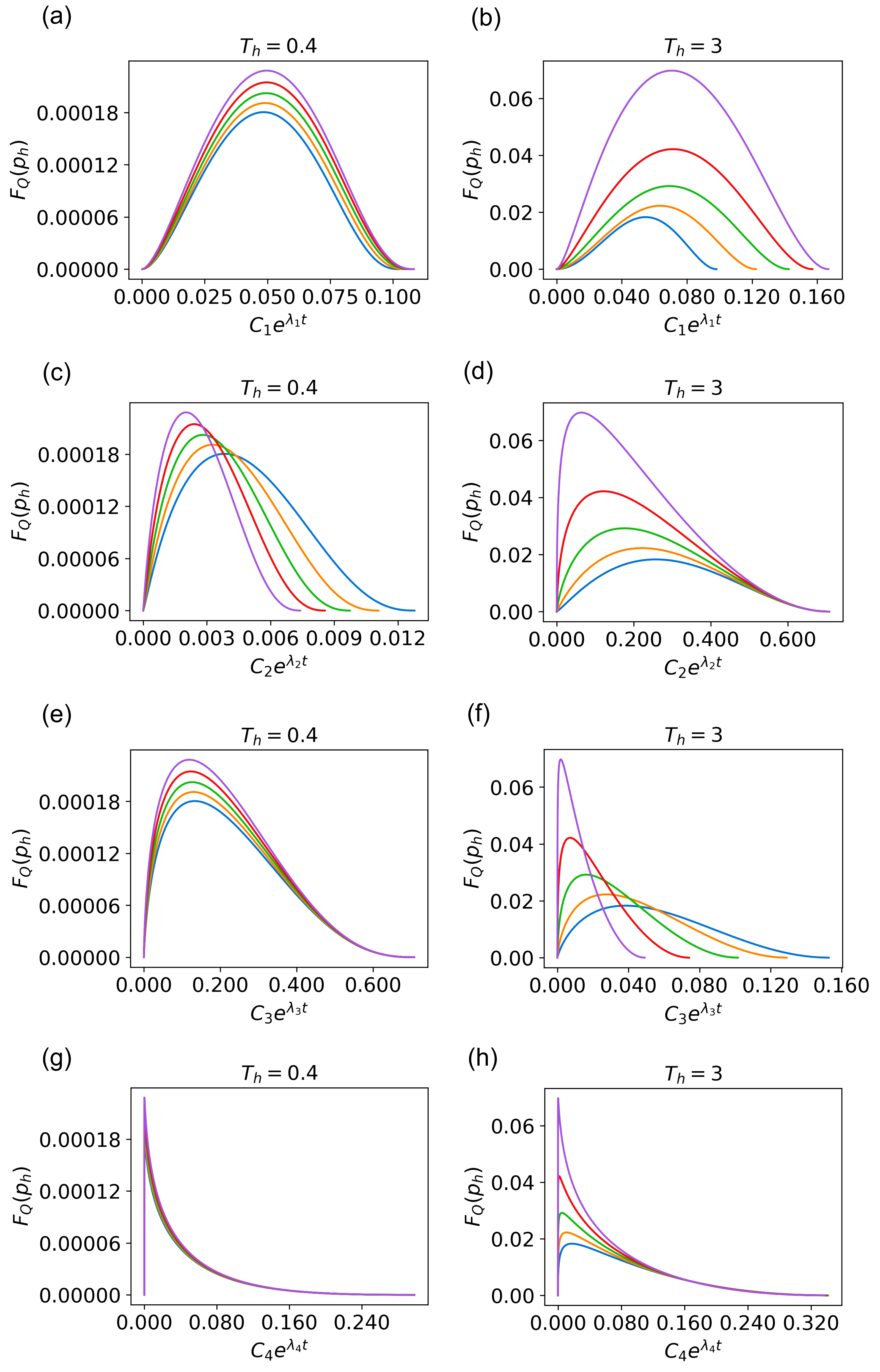}
\caption{$F_Q(p_h)$ as a function of the weighted Liouvillian mode contribution $c_n e^{\lambda_n t}$ for the estimation of different hot-bath coherence $p_h$ with fixed $p_c$ = 0.7. (a,b), (c,d), (e,f), and (g,h) correspond to modes $n=1,2,3,$ and $4$ respectively.}
\label{F_ph_modal}
\end{figure} 
    
To elucidate the above observations, we analyze the expansion coefficients $c_n$ ($n=0,1,2,3,4$) appearing in Eq.~(\ref{eq:cn}), which quantify the projection of the initial state onto the corresponding Liouvillian eigenmodes. Since $c_0$ is associated with the stationary eigenmode ($\lambda_0=0$), it contributes only to the steady-state density matrix and is therefore not relevant to the transient dynamics considered here. The remaining coefficients characterize the statistical importance of the transient relaxation modes. Their dependence on the hot-bath coherence parameter $p_h$ is shown in Fig.~(\ref{F_ph_C_n}e,f). At both low and high hot-bath temperatures, the magnitudes of the four transient coefficients are comparable. At low temperature, $c_2<c_1<c_4<c_3$ and these remain nearly independent of $p_h$. At higher temperature, however, a crossover occurs between the $n=1$ and $n=3$ modes, with $c_1$ and $c_3$ being very close to each other over the range of $p_h$ considered.

Although the comparable magnitudes of the coefficients suggest that all relaxation modes participate in the dynamics, the transient metrological response is not determined by the coefficients alone. Rather, it is governed by the time-dependent modal amplitudes $c_ne^{\lambda_nt}$, where $\lambda_n$ are the Liouvillian eigenvalues. These quantities incorporate both the statistical importance of a mode through $c_n$ and its dynamical evolution through the exponential relaxation factor. To identify which relaxation channels are primarily responsible for the transient enhancement of $F_Q(p_h)$, we therefore examine the QFI as a function of the weighted modal amplitudes $c_ne^{\lambda_nt}$, as shown in Fig.~(\ref{F_ph_modal}a-h) for $T_h=0.4$ and $3$. Each subplot contains five curves corresponding to different values of the coherence parameter $p_h$. With the exception of the fourth relaxation mode ($n=4$), all weighted modal amplitudes exhibit well-defined transient maxima whose magnitude and location depend on $p_h$.

Fig.~(\ref{F_ph_modal}a,b) show the dependence of the QFI for estimating $p_h$ on the weighted amplitude $c_1e^{\lambda_1t}$ at low and high temperatures, respectively. In both cases the QFI exhibits a transient maximum. At lower temperature Fig.~(\ref{F_ph_modal}a), increasing $p_h$ primarily suppresses the magnitude of the peak while leaving its position almost unchanged. At higher temperature Fig.~(\ref{F_ph_modal}b), both the peak height and its position vary with $p_h$, with the peak shifting toward shorter scaled times as its magnitude decreases.

The dependence of the QFI on the weighted amplitude $c_2e^{\lambda_2t}$ is shown in Fig.~(\ref{F_ph_modal}c,d). At both temperatures, the QFI peak shifts systematically toward lower values of $c_2e^{\lambda_2t}$ as $p_h$ is varied. At lower temperature, $c_2$ is the smallest among the four transient coefficients Fig.~(\ref{F_ph_C_n}e), while the relaxation rates satisfy
$
|\lambda_1|=\Delta<|\lambda_2|<|\lambda_3|<|\lambda_4|$.
Despite the fact that $\lambda_2$ corresponds to a faster relaxation channel than the Liouvillian-gap mode, the corresponding scaled time window in Fig.~(\ref{F_ph_modal}c) is compressed by more than a factor of ten compared with those of the other modes shown in Fig.~(\ref{F_ph_modal}a,e,g). This compression originates from the combined influence of the modal coefficient $c_2$ and the exponential relaxation factor $e^{\lambda_2t}$. Consequently, although the slowest relaxation timescale is determined by $\lambda_1=\Delta$, it is the $n=2$ relaxation channel that provides the dominant contribution to the emergence of the transient optimum at lower temperatures. This demonstrates that the appearance of transient metrological enhancement is not governed solely by the slowest relaxation mode but by the competition between the statistical importance of a mode, encoded in $c_n$, and its relaxation dynamics, determined by $\lambda_n$.

The resulting nonlinear dependence of $t^*$ on $1/\Delta$ therefore does not arise simply from the participation of multiple relaxation modes. Rather, it originates from the competition between the Liouvillian-gap mode, which sets the slowest relaxation timescale, and a faster mode whose statistical weight is sufficient to dominate the transient enhancement. The emergence of nonlinear scaling is therefore not, by itself, an indicator of genuine multimode dynamics.

The converse situation is observed at higher temperatures. As shown in Fig.~(\ref{F_ph_C_n}f), the coefficients $c_1$ and $c_3$ become comparable in magnitude, while the relaxation rates continue to satisfy
$
|\lambda_1|=\Delta<|\lambda_2|<|\lambda_3|<|\lambda_4|.
$
Consequently, both the $n=1$ and $n=3$ weighted modal amplitudes produce strongly compressed scaled time windows, as observed in Fig.~(\ref{F_ph_modal}b,f), indicating that both relaxation channels contribute significantly to the transient enhancement of the QFI despite their substantially different relaxation rates. Since the corresponding scaled time windows occur within the same order of magnitude, the remaining two modes also contribute, although less prominently than the $n=1$ and $n=3$ channels. This regime therefore represents genuine multimode transient dynamics. Nevertheless, Fig.~(\ref{F_ph_C_n}d) shows that the optimal interrogation time continues to satisfy the approximate inverse-gap scaling relation $t^*\propto1/\Delta$. The present results therefore establish that neither nonlinear deviations from inverse-gap scaling nor the persistence of linear scaling can, by themselves, be regarded as signatures of multimode dynamics. Instead, the transient metrological behavior is determined by the interplay between the statistical importance of the relaxation modes and their characteristic relaxation timescales.

\begin{figure}
\centering
    \includegraphics[width=1\linewidth]{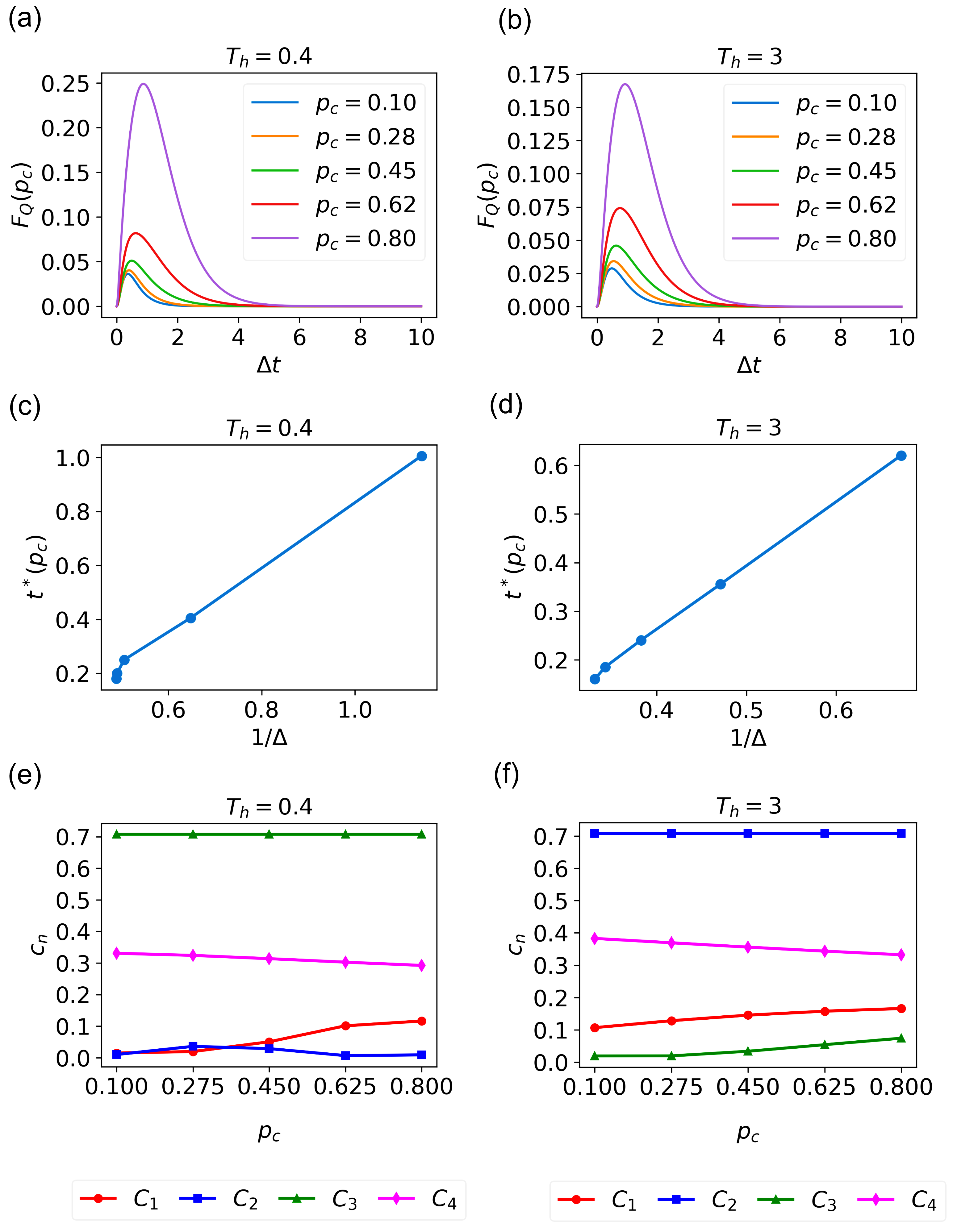}
\caption{(a,b) $F_Q(p_c)$ as a function of rescaled time $\Delta t$. (c,d) Optimal estimation time $t^*(p_c)$ as a function of the inverse of Liouvillian gap $\Delta$. (e,f) $c_n$ as a function of $p_c$.}

\label{F_pc_C_n}
\end{figure}

We now consider the estimation of the cold-bath coherence parameter, $p_c$. In Fig.~(\ref{F_pc_C_n}a,b), we show the time evolution of $F_Q(p_c)$ for $T_h=0.4$ and $3$, respectively. Similar to the hot-bath coherence parameter, $F_Q(p_c)$  exhibits a pronounced transient maximum before decaying to zero at steady state, indicating that the optimal estimation of $p_c$ is likewise achieved during the transient dynamics. Increasing the hot-bath temperature decreases the magnitude of the transient QFI, implying that lower temperatures are more favorable for estimating the cold-bath coherence parameter.

The corresponding optimal interrogation time $t^*$ is shown as a function of the inverse Liouvillian gap in Fig.~(\ref{F_pc_C_n}c,d). As in the case of $p_c$, the dependence is approximately linear at both low and high temperatures. However, as established in the previous section, this linearity should not be interpreted as evidence of Liouvillian-gap dynamics. The modal expansion coefficients $c_n$, shown in Fig.~(\ref{F_pc_C_n}e,f), remain comparable in magnitude over the entire range of $p_c$. At lower temperatures Fig.~(\ref{F_pc_C_n}e), $c_1\approx c_2$ for small values of $p_c$, with the two coefficients crossing such that $c_1>c_2$ as $p_c$ increases. At higher temperatures Fig.~(\ref{F_pc_C_n}f), the coefficients satisfy $c_3<c_1<c_4<c_2$, although they remain of comparable magnitude throughout the parameter range. Their weak dependence on $p_c$ indicates that varying the cold-bath coherence does not significantly alter the statistical importance of the Liouvillian relaxation modes. Consequently, the qualitative features of the transient estimation dynamics for $p_c$ closely resemble those obtained for $p_h$.
%%%%%%%%%%%%%%%%%%%

%%%%%%%%%%%%%%
%%%%%%%%%%%%%%%%%%%%%%%%%%%%%%%%%%%%%%%%%%%%%
\begin{figure}
\centering
    \includegraphics[width=1\linewidth]{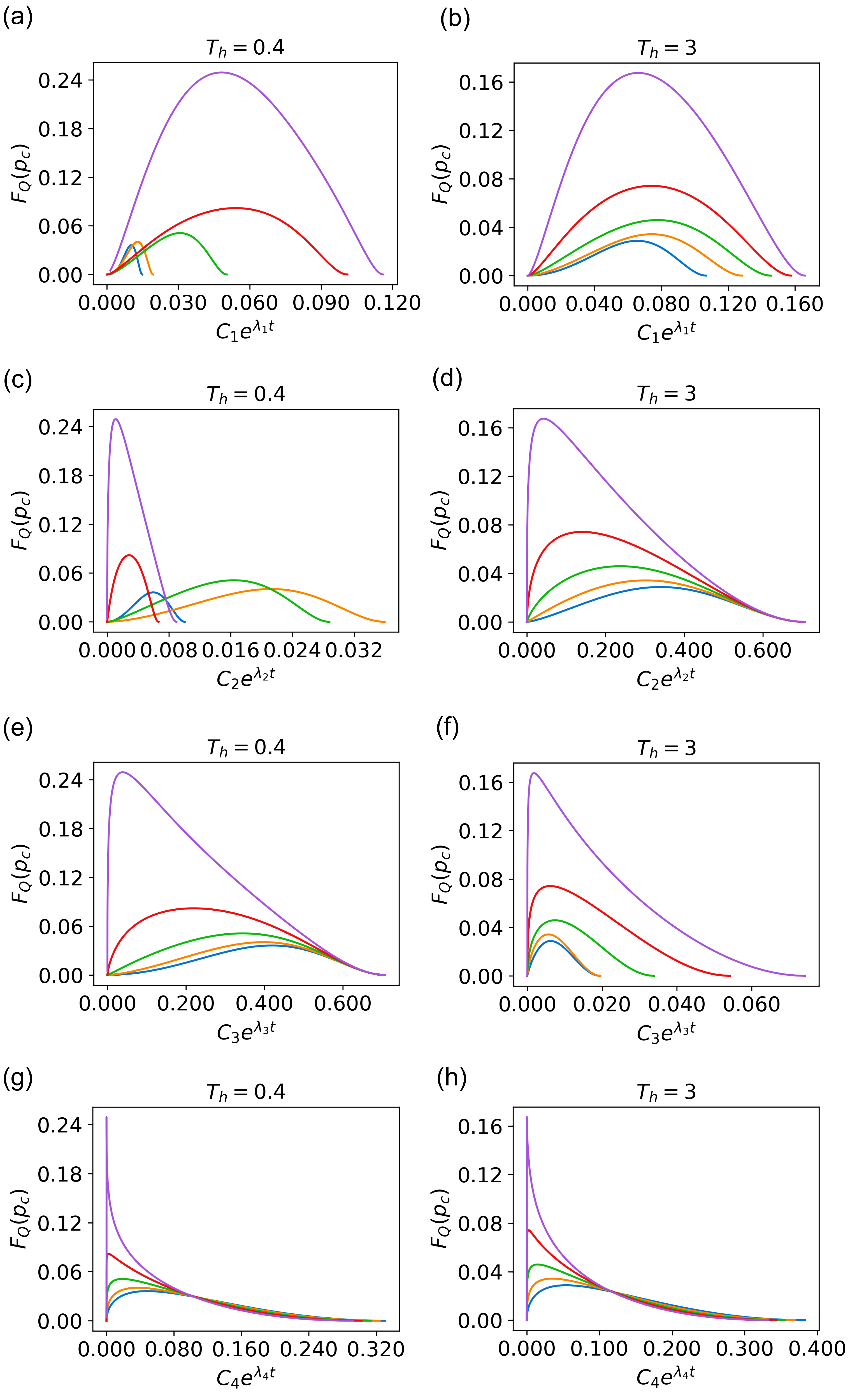}
\caption{$F_Q(p_c)$ as a function of the weighted Liouvillian mode contribution $c_n e^{\lambda_n t}$. (a,b), (c,d), (e,f), and (g,h) correspond to modes $n=1,2,3,$ and $4$ respectively.}
\label{F_pc_modal}
\end{figure} 
%%%%%%%%%%%%%%%%%%%%
Further insight is obtained by examining the dependence of the QFI on the weighted modal amplitudes $c_ne^{\lambda_nt}$, shown in Fig.~(\ref{F_pc_modal}a-h). At lower temperatures, the largest compression of the scaled time window is observed for the $n=1$ and $n=2$ modes at small values of $p_c$. Comparing the blue curves corresponding to $p_c=0.1$ in Fig.~(\ref{F_pc_modal}a) and Fig.~(\ref{F_pc_modal}c) shows that both modes possess nearly identical compressed intervals, extending approximately from $0$ to $0.01$. In the same parameter regime, Fig.~(\ref{F_pc_C_n}e) shows that $c_1\approx c_2$. The comparable statistical importance of these modes, together with the similar compression of their scaled time windows, indicates that both relaxation channels contribute nearly equally to the transient enhancement of the QFI. Thus, despite the simultaneous participation of two dominant modes, the optimal interrogation time continues to satisfy the approximate inverse-gap relation $t^*\propto1/\Delta$. This again demonstrates that linear scaling of the optimal interrogation time should not even be regarded as a definitive signature of single-mode dynamics.

At higher temperatures, a different scenario emerges. As shown in Fig.~(\ref{F_pc_C_n}f), the $n=3$ mode possesses the smallest expansion coefficient, yet its weighted modal amplitude exhibits the largest compression of the scaled time window among all four transient modes. The competition between the statistical importance of the relaxation modes and their respective relaxation rates therefore favors the $n=3$ channel in determining the transient enhancement, even though it is not associated with the Liouvillian gap. Nevertheless, the optimal interrogation time continues to obey the approximate scaling relation $t^*\propto1/\Delta$. These observations further support the conclusion that the persistence of inverse-gap scaling does not necessarily imply that the transient dynamics are governed by the Liouvillian gap mode. Rather, the transient metrological response is determined by the interplay between the statistical importance of the relaxation modes and their characteristic relaxation timescales, with faster relaxation channels capable of dominating the transient enhancement when their effective weighted modal amplitudes become sufficiently compressed.

Throughout the parameter regime considered in this work, we find that among the four non-steady Liouvillian modes there always exists one mode whose expansion coefficient satisfies $
c_m=\frac{1}{\sqrt{2}},
$
where the mode index $m$ keeps changing as system parameters change. In Fig.~(\ref{F_ph_C_n}e,f) and Fig.~(\ref{F_pc_C_n}e,f), one of the four modes is always as 0.707.  The invariant quantity is not the projection onto a particular Liouvillian eigenmode, but rather the existence of a relaxation mode carrying a fixed modal weight. The identity of this mode interchanges with the remaining eigenmodes as the Liouvillian spectrum evolves. The existence of a mode with the fixed amplitude
$c_m=1/\sqrt2$ provides a parameter-independent reference weight, but this mode does not necessarily coincide with the slowest relaxation mode associated with the Liouvillian gap.
 
\begin{figure}
\centering
    \includegraphics[width=1\linewidth]{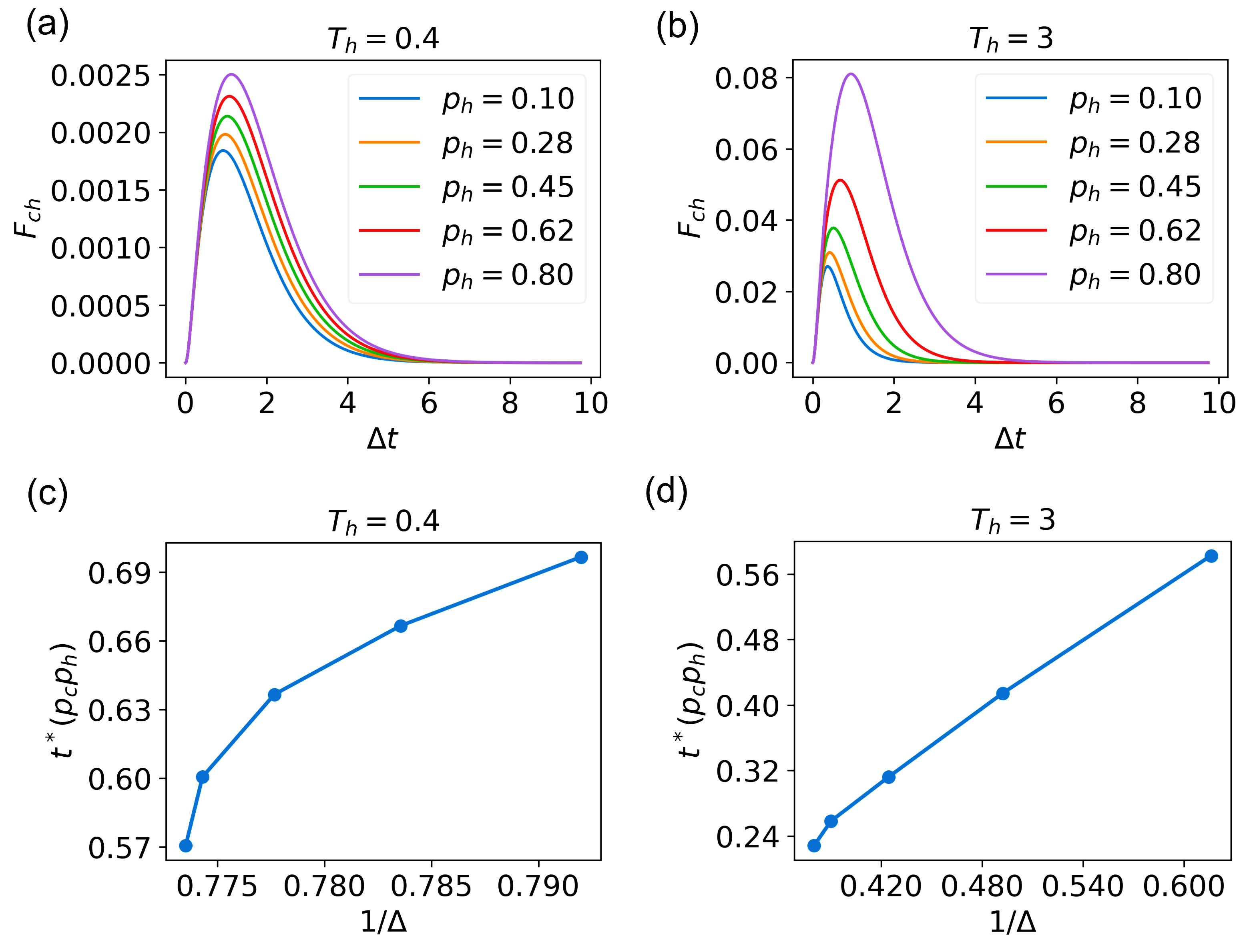}
\caption{Simultaneous estimation of hot-bath coherence parameter $p_h$ and cold-bath coherence parameter $p_c$. (a,b) $F_{ch}$ as a function of rescaled time $\Delta t$. (c,d) Optimal estimation time $t^*(p_cp_h)$ as a function of the inverse of Liouvillian gap $\Delta$. In all panels $p_h$ is varied while $p_c$ is fixed at $p_c$ = 0.7.}

\label{fig-F_ch_delta}
\end{figure}

We next examine the off-diagonal element of the QFIM, $F_{ch}$, which quantifies the statistical correlation between the simultaneous estimation of the hot- and cold-bath coherence parameters. The time evolution of $F_{ch}$ is shown in Fig.~(\ref{fig-F_ch_delta}a,b) for $T_h=0.4$ and $3$, respectively. Similar to the diagonal QFI elements, $F_{ch}$ exhibits a pronounced transient maximum before vanishing in the steady state, indicating that the statistical correlations between the two estimation parameters are likewise strongest during the transient dynamics. The magnitude of the transient peak increases with increasing hot-bath temperature.

\begin{figure}
\centering
    \includegraphics[width=1\linewidth]{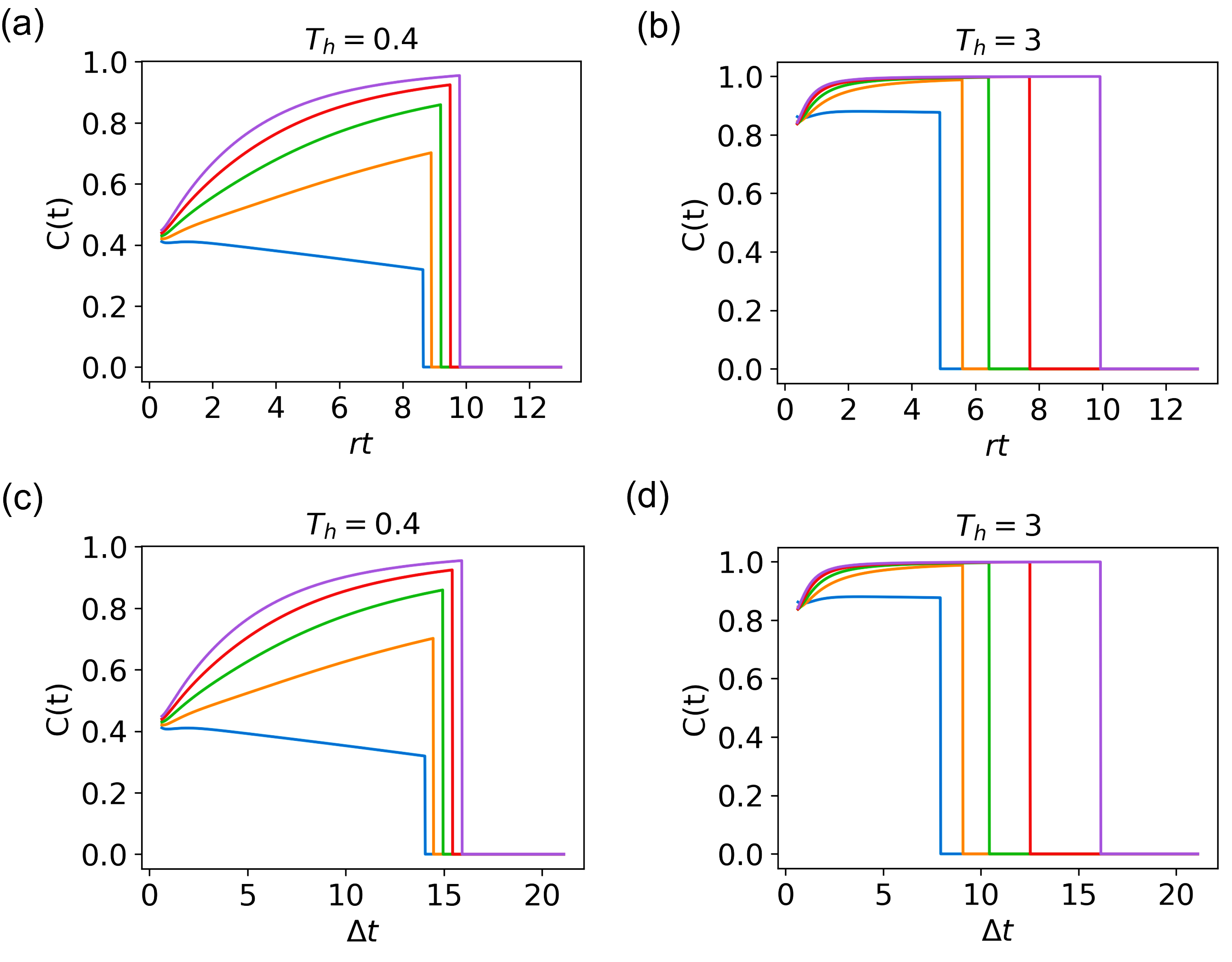}
\caption{ QFIM correlation coefficient, $C(t)$ for simultaneous estimation of the hot- and cold-bath coherence parameters. (a,b) $C(t)$ as a function of time. (c,d) $C(t)$ as a function of rescaled timescale $\Delta t$. In all panels $p_h$ is varied and $p_c$ is fixed at 0.7.}

\label{Correlation}
\end{figure}

The corresponding optimal interrogation time, obtained from the maximum of $F_{ch}$, is shown as a function of the inverse Liouvillian gap in Fig.~(\ref{fig-F_ch_delta}c,d). A pronounced nonlinear dependence is observed at the lower temperature, whereas an approximately linear trend emerges at the higher temperature. Consistent with the diagonal elements of the QFIM, the observed scaling of the optimal interrogation time alone does not distinguish between single- and multimode transient dynamics. Rather, it reinforces the conclusion that the transient metrological behavior is governed by the collective influence of the Liouvillian relaxation modes. To further quantify the interplay between the simultaneous estimation of the hot- and cold-bath coherence parameters, we next examine the normalized correlation coefficient.
\begin{equation}
C(t)=\frac{F_{ch}}{\sqrt{F_{cc}F_{hh}}},
\end{equation}
which provides a dimensionless measure of the statistical coupling between the two estimators. Fig.~(\ref{Correlation}a,b) show $C(t)$ as a function of time for low and high hot-bath temperatures, while Fig.~(\ref{Correlation}c,d) present the same quantity as a function of the scaled time $\Delta t$. Each subplot contains five curves corresponding to different values of $p_h$.

At both temperatures, all curves originate from the same initial value and subsequently increase at rates that depend on $p_h$. Throughout the evolution, $C(t)$ remains below unity, indicating that the statistical coupling between the estimators remains finite without reaching the limit of maximal correlation. As the system approaches the steady state, each curve undergoes a rapid collapse to zero, with the collapse occurring at different times for different values of $p_h$. This behavior reflects the parameter dependence of the transient estimation window: while the normalized statistical coupling builds up during the nonequilibrium evolution, it vanishes once the corresponding QFI elements decay to zero. The variation in the collapse time further demonstrates that the duration over which the two parameters remain jointly informative is controlled by the coherence parameter, even though the overall qualitative behavior is preserved across the entire parameter range.
The ability to estimate the two coherence parameters simultaneously is inherently transient. During the nonequilibrium evolution, the statistical coupling between the estimators builds up, indicating that the sensitivities to $p_h$ and $p_c$
 become increasingly interdependent. However, because the normalized coupling remains below unity, the information carried by the two parameters never becomes completely redundant, allowing them to remain distinguishable throughout the transient regime. As the system relaxes, the statistical coupling disappears together with the QFI, limiting simultaneous estimation to a finite temporal window whose duration depends on the coherence parameter.
\section{Transient to Steadystate Estimation}

We now proceed to show that the transient nature of the metrological advantage can be actively controlled. We now investigate whether the transient metrological behavior can be controlled by coherently coupling the two upper states of the V-type system to a single-mode cavity with matter-field coupling strength $g$.  In this case the total Hamiltonian of the system gets two additional terms, $\hat H = \hat H_T +\hat H_\ell+\hat H_{s\ell}$. 
The cavity Hamiltonian is
$
\hat H_\ell=\hbar\omega_\ell\,\hat a_\ell^\dagger\hat a_\ell,
$ and the system–cavity coupling Hamiltonian  is  
$
\hat H_{s\ell}=g\big(\hat a_\ell\,\hat c_b^\dagger+\hat a_\ell^\dagger\,\hat c_b\big),$ with \(g\) is the coupling strength. Under the same Born-Markov-secular regime by treating the system-cavity Hamiltonian perturbatively, the reduced density matrix elements remain the same and the superoperator Liouvillian  changes to $\mathcal L_{\ell}={\cal L}+{\cal L}_{cav}$, with
\begin{equation}\label{eq:Lmatrix}
\mathcal L_{cav} =
\begin{pmatrix}
0& 0 & 0 & 0 & 0\\[2pt]
0 & 0 &0 &0 & 0\\[2pt]
0 &0 & -g^2(1+ n_\ell) & g^2 n_\ell & 0\\[2pt]
0 & 0 & g^2(1+ n_\ell) & -g^2 n_\ell & 0\\[2pt]
0 & 0 & 0 &0 & 0
\end{pmatrix}.
\end{equation}
where $n_\ell$ is the cavity's Bose-Einstein function. We focus on the regime $g>r$, where the coherent system-cavity interaction exceeds the system-bath coupling. For completeness, we note that when $g<r$, the cavity does not qualitatively modify the dynamics, and the QFI retains the transient behavior discussed previously.

\begin{figure}
\centering
    \includegraphics[width=1\linewidth]{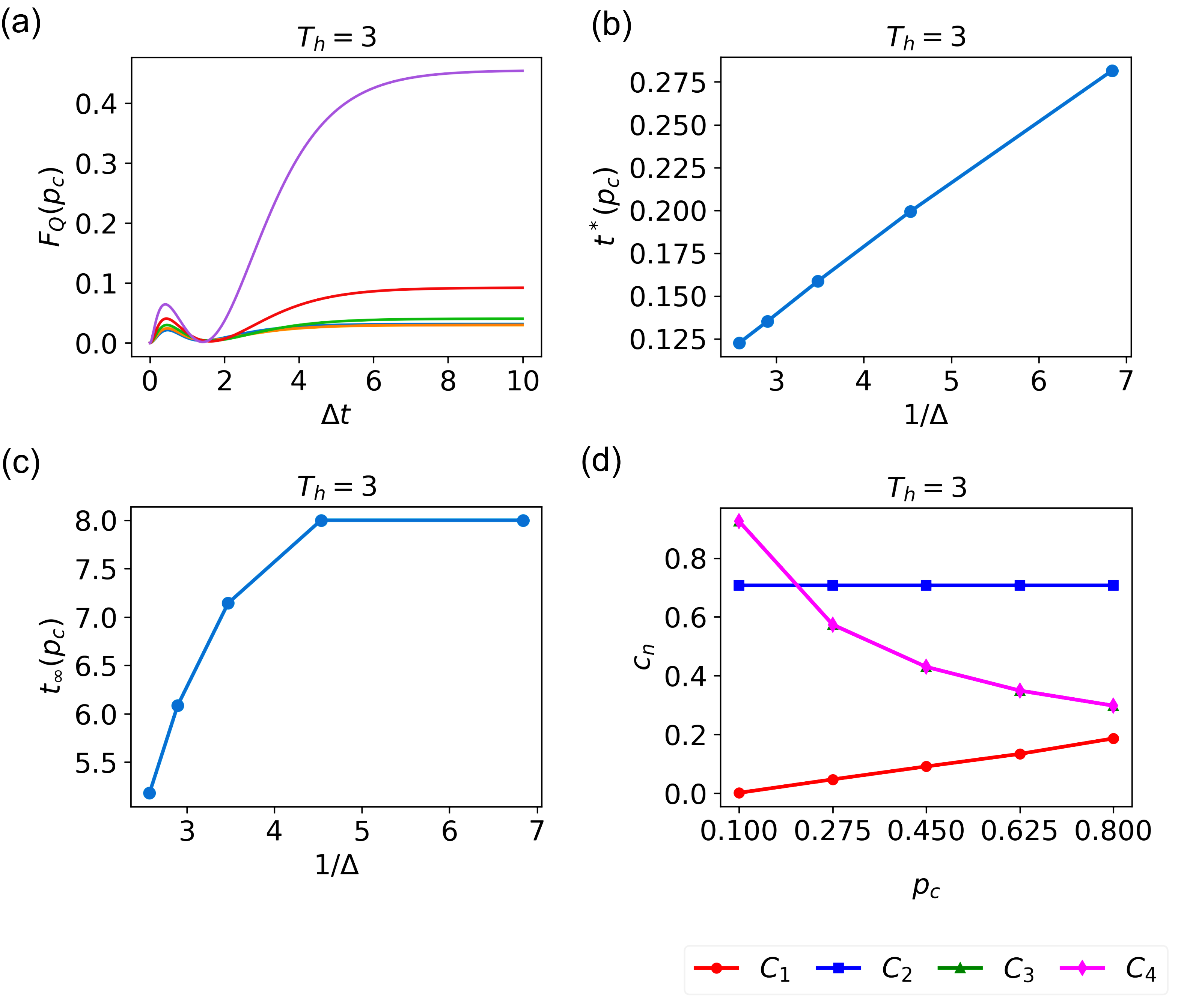}
\caption{(a) $F_Q(p_c)$ as a function of rescaled time $\Delta t$ for a cavity coupled system with $r=0.1, g =1$. (b) Optimal estimation time $t^*(p_c)$ as a function of the inverse of Liouvillian gap  $\Delta$ for the early $F_Q(p_c)$ peak. (c) saturation time $t_\infty$ as a function of the inverse of Liouvillian gap $\Delta$. (d) $c_n$ as a function of $p_c$.}

\label{F_pc_g_greater_r}
\end{figure}

In Fig.~(\ref{F_pc_g_greater_r}a,b), we show the evolution of $F_Q(p_c)$, using ${\cal L}_\ell$ is the effective evolution operator at low and high hot-bath temperatures, respectively. At the lower temperature, the QFI continues to exhibit a transient maximum followed by decay, indicating that coherent cavity coupling alone is insufficient to stabilize the metrological sensitivity. In contrast, at the higher temperature, $F_Q(p_c)$ displays a small transient overshoot before saturating to a finite steady-state value. Thus, only when coherent system-cavity coupling is accompanied by sufficiently strong thermal excitation does the transient metrological advantage evolve into a sustained steady-state resource.

To characterize the onset of this steady regime, we extract the saturation time $t_\infty$, shown as a function of the inverse Liouvillian gap in Fig.~(\ref{F_pc_g_greater_r}c). The observed nonlinear dependence indicates that the establishment of the steady metrological regime is not governed solely by the slowest relaxation timescale, but instead results from the collective dynamics of multiple Liouvillian eigenmodes.

Further insight is obtained from the modal expansion coefficients shown in Fig.~(\ref{F_pc_g_greater_r}d). While the modal weights remain comparable in magnitude, they now exhibit a noticeable dependence on the coherence parameter $p_c$, in contrast to the cavity-free case where they were nearly parameter independent. This suggests that the cavity modifies how the Liouvillian eigenmodes participate in the dynamics, allowing the parameter dependence to be redistributed among the modes. The emergence of a finite steady-state QFI at higher temperature therefore appears to arise from the combined action of coherent cavity coupling and thermal driving, rather than from coherent control alone.

Taken together, these results suggest that transient and steady-state quantum metrology correspond to two distinct dynamical regimes of the same open quantum system. In the absence of sufficiently strong coherent control, parameter sensitivity is generated and lost during the transient evolution, making the optimal estimation time a dynamical quantity determined by the interplay of several Liouvillian modes. When coherent system-cavity coupling dominates and is assisted by thermal excitation, this transient metrological advantage evolves into a sustained steady-state resource. The cavity therefore acts not simply as an additional dissipative channel, but as a control element that modifies the dynamical balance between coherent evolution and irreversible relaxation, enabling long-lived parameter sensitivity under appropriate operating conditions.

%%%%%%%%%%%%%%%

%%%%%%%%%%%%%%%%%%%%%%%%%%%%%%%%%%%%%%%%%%%%%%%

%%%%%%%%%%%%%%%
%%%%%%%%%%%%%%

%%%%%%%%%%

\section{Analytical Theory of QFI Decomposition }

%\subsection{Multimode Liouvillian parameter encoding and transient quantum metrology}

%%%%%%%%%%%%%%SELF%%%%%%%%%%%%%%
%%%%%%%%%%%%%%%%%%%%%%%%%%%%%%%%%

Differentiating Eq.~(\ref{eq:rho_spec}) with respect to a estimation parameter, $\theta_\mu$ gives
\begin{align}
\partial_{\theta_\mu}\rho
=
&
\sum_{n>0}
e^{\lambda_nt}
\left(
\partial_{\theta_\mu}c_n
+
tc_n
\partial_{\theta_\mu}\lambda_n
\right)
r_n
\nonumber\\
&
+
\sum_{n>0}
c_n
e^{\lambda_nt}
\partial_{\theta_\mu}r_n.
\label{eq:drho_modes}
\end{align}
Eq.~(\ref{eq:drho_modes}) shows that the parameter dependence of the density operator arises through three distinct mechanisms: the variation of the modal amplitudes, the parameter dependence of the Liouvillian eigenvalues, and the deformation of the relaxation eigenoperators. While it is well-understood that every element of the QFIM receives contributions from all modes, not much can be told about the quantitative behavior of the transient QFIM with respect to the parameters encoded in the eigenmodes directly. We next proceed to show that some interesting analytical inferences can sill be drawn if we transform the QFIM to the Liouvillian's eigenspace.
Differentiating the formal solution of the master equation with respect to the parameter $\theta_\mu$ (the dependence on $\theta$ is suppressed for brevity) gives
\begin{equation}
\partial_\mu\rho(t)
=
\left(
\partial_\mu e^{\mathcal Lt}
\right)\rho_0.
\end{equation}
The Fr\'echet derivative of the exponential map is
\begin{equation}
\partial_\mu
e^{\mathcal Lt}
=
\int_0^t
e^{\mathcal L(t-s)}
\left(
\partial_\mu\mathcal L
\right)
e^{\mathcal Ls}
\,ds.
\label{eq:frechet}
\end{equation}

For a diagonalizable Liouvillian, the spectral decomposition is
\begin{equation}
\mathcal L
=
\sum_n
\lambda_n
P_n,
\label{eq:spectral}
\end{equation}
where the spectral projectors satisfy
$
P_mP_n=\delta_{mn}P_n
$
and
$
\sum_nP_n=I.
$
The propagator therefore becomes
\[
e^{\mathcal Lt}
=
\sum_n
e^{\lambda_nt}
P_n.
\]

Substituting Eq.~(\ref{eq:spectral}) into Eq.~(\ref{eq:frechet}) yields
\begin{equation}
\partial_\mu\rho(t)
=
\sum_{m,n}
f_{mn}(t)
P_m
(\partial_\mu\mathcal L)
P_n
\rho_0,
\label{eq:drho}
\end{equation}
where
\begin{equation}
f_{mn}(t)
=
\begin{cases}
\dfrac{
e^{\lambda_mt}
-
e^{\lambda_nt}
}
{\lambda_m-\lambda_n},
&
\lambda_m\neq\lambda_n,
\\[2ex]
t\,e^{\lambda_mt},
&
\lambda_m=\lambda_n.
\end{cases}
\label{eq:kernel}
\end{equation}
The functions $f_{mn}(t)$ constitute the temporal kernels associated with the Liouvillian spectrum. Applying the vectorization map to Eq.~(\ref{eq:drho}) and using the linearity of the $\mathrm{vec}(\cdot)$ operation gives
\begin{equation}
\mathrm{vec}(\partial_\mu\rho)
=
\sum_{m,n}
f_{mn}(t)
\,
\mathrm{vec}
\!\left[
P_m
(\partial_\mu\mathcal L)
P_n
\rho_0
\right].
\end{equation}
It is therefore convenient to define the modal responses
\begin{equation}
v_{mn}^{(\mu)}
=
\mathrm{vec}
\!\left[
P_m
(\partial_\mu\mathcal L)
P_n
\rho_0
\right],
\label{eq:vmu}
\end{equation}
so that
\begin{equation}
\mathrm{vec}(\partial_\mu\rho)
=
\sum_{m,n}
f_{mn}(t)
v_{mn}^{(\mu)}.
\label{eq:drho_vec}
\end{equation}
Using the vectorization identity
\[
\mathrm{vec}(AXB)
=
(B^T\otimes A)\,
\mathrm{vec}(X),
\]
the Sylvester equation defining the symmetric logarithmic derivative becomes
\begin{equation}
\mathrm{vec}(\partial_\mu\rho)
=
\frac12
\left(
I\otimes\rho
+
\rho^T\otimes I
\right)
\mathrm{vec}(L_\mu).
\end{equation}
Introducing the Sylvester superoperator
\begin{equation}
S_\rho
=
I\otimes\rho
+
\rho^T\otimes I,
\end{equation}
and assuming that $\rho(t)$ is full rank (otherwise $S_\rho^{-1}$ is replaced by its Moore-Penrose pseudoinverse), one obtains
\begin{equation}
\mathrm{vec}(L_\mu)
=
2
S_\rho^{-1}
\,
\mathrm{vec}(\partial_\mu\rho).
\label{eq:vectorSLD}
\end{equation}
Substituting Eq.~(\ref{eq:drho_vec}) into Eq.~(\ref{eq:vectorSLD}) gives
\begin{equation}
\mathrm{vec}(L_\mu)
=
2
S_\rho^{-1}
\sum_{m,n}
f_{mn}(t)
v_{mn}^{(\mu)}.
\label{eq:Lvector}
\end{equation}
The quantum Fisher information matrix may be written as
\begin{equation}
F_{\mu\nu}\!
=\!
\mathrm{Re}
\!\left[
\mathrm{vec}(L_\mu)^\dagger\!
(\rho^T\otimes I)\!
\mathrm{vec}(L_\nu)
\right].
\label{eq:qfi_vec}
\end{equation}
Substituting Eq.~(\ref{eq:Lvector}) into Eq.~(\ref{eq:qfi_vec}) yields
\begin{align}
F_{\mu\nu}\!
=\!
4\!\mathrm{Re}
\Bigg[\!
&
\left(
\sum_{m,n}
f_{mn}(t)
v_{mn}^{(\mu)}
\right)^\dagger\!
S_\rho^{-1}
(\rho^T\!\otimes\! I)
S_\rho^{-1}\!
\left(\!
\sum_{k,l}
f_{kl}(t)
v_{kl}^{(\nu)}\!
\right)\!
\Bigg].
\end{align}
Expanding the quadratic form over the modal indices gives
\begin{align}
F_{\mu\nu}
=\!
4\,\mathrm{Re}
\Bigg[
\sum_{m,n}
\sum_{k,l}
&
f_{mn}(t)
f_{kl}(t)
\nonumber\\
&
\times
\left(v_{mn}^{(\mu)}\right)^\dagger
S_\rho^{-1}
(\rho^T\otimes I)
S_\rho^{-1}
v_{kl}^{(\nu)}
\Bigg],
\label{eq:qfimfinal}
\end{align}
which provides an exact decomposition of the QFIM in terms of Liouvillian spectral projectors. The explicit dependence on the Liouvillian spectrum is contained in the temporal kernels $f_{mn}(t)$, while the instantaneous quantum statistical geometry is encoded in the metric
$
S_\rho^{-1}
(\rho^T\otimes I)
S_\rho^{-1}.
$ Since $\rho=\rho(t)$, both $S_\rho$ and $\mathcal M$ are generally time dependent.
For convenience, the modal coupling coefficients are defined as
\begin{equation}
A_{mnkl}^{\mu\nu}(t)
=
4\,\mathrm{Re}
\!\left[
\left(v_{mn}^{(\mu)}\right)^\dagger
S_\rho^{-1}
(\rho^T\otimes I)
S_\rho^{-1}
v_{kl}^{(\nu)}
\right],
\label{eq:Acoeff}
\end{equation}
where the time dependence arises through the instantaneous density operator entering the Sylvester superoperator. The object
$A_{mnkl}^{\mu\nu}(t)$ contains the modal response vectors,
the Sylvester metric and 
the instantaneous density matrix and is hence much richer than a simple weight. It is actually a coupling tensor between pairs of Liouvillian modes, i.e a 
modal coupling tensor. Equation~(\ref{eq:qfimfinal}) can now be written compactly as
\begin{equation}
F_{\mu\nu}(t)
=
\sum_{m,n}
\sum_{k,l}
A_{mnkl}^{\mu\nu}(t)
f_{mn}(t)
f_{kl}(t).
\label{eq:qfim_modal}
\end{equation}
 Equation~(\ref{eq:qfim_modal}) therefore separates the QFIM into universal temporal kernels determined by the Liouvillian spectrum (a spectral contribution) and modal coupling tensors (statistical contribution) that encode the parameter sensitivity of the dynamics together with the instantaneous quantum statistical metric.
%%%%%%%%%%%%%%%%%%
The optimal interrogation time is determined by the stationarity condition
$
\dot F_{\mu\nu}(t)=0.
$
Differentiating Eq.~(\ref{eq:qfim_modal}) yields
\begin{align}
\frac{dF_{\mu\nu}}{dt}
=
\sum_{m,n}
\sum_{k,l}
&
\dot A^{\mu\nu}_{mnkl}(t)
f_{mn}(t)
f_{kl}(t)
\nonumber\\
&+
A^{\mu\nu}_{mnkl}(t)
\left[
\dot f_{mn}(t)
f_{kl}(t)
+
f_{mn}(t)
\dot f_{kl}(t)
\right],
\label{eq:stationary_general}
\end{align}
The first term originates from the time dependence of the quantum statistical metric through $\rho(t)$, while the second describes the evolution of the temporal kernels determined by the Liouvillian spectrum. Equation~(\ref{eq:stationary_general}) therefore constitutes the exact condition determining the optimal interrogation time.

The expression simplifies considerably once the density matrix approaches the stationary state. Since $\rho(t)\rightarrow\rho_{\mathrm{ss}}$, the Sylvester superoperator $S_\rho$ and the metric $\mathcal M$ become time independent, implying $
\dot A^{\mu\nu}_{mnkl}(t)\rightarrow0.
$ The stationarity condition then reduces to
\begin{align}
\frac{dF_{\mu\nu}}{dt}
=
\sum_{m,n}
\sum_{k,l}
A^{\mu\nu}_{mnkl}
\left[
\dot f_{mn}(t)
f_{kl}(t)
+
f_{mn}(t)
\dot f_{kl}(t)
\right],
\label{eq:stationary_ss}
\end{align}
where the coefficients are now evaluated using the steady-state density matrix.
% Considering the limiting case in which a single Liouvillian mode dominates the parameter encoding, one recovers the conventional inverse-gap scaling. In this limit,
% $
% A^{\mu\nu}_{1111}
% \gg
% A^{\mu\nu}_{mnkl},
% \qquad
% (m,n,k,l)\neq(1,1,1,1),
% $
% so that
% \begin{equation}
% F_{\mu\nu}(t)
% \simeq
% A^{\mu\nu}_{1111}
% t^2
% e^{-2\Delta t},
% \label{eq:singlemode}
% \end{equation}
% where $\Delta=-\mathrm{Re}(\lambda_1)$ is the Liouvillian gap. The stationarity condition becomes
% $
% d_t
% \left(
% t^2
% e^{-2\Delta t}
% \right)
% =
% 0,$
% which gives
% $
% t_{\mathrm{opt}}
% =
% 1/\Delta.
% $
% up to an overall numerical prefactor determined by the precise definition of the characteristic interrogation time.
% It  represents the asymptotic single-mode limit of the general condition (\ref{eq:stationary_general}). In the steady-state regime, where the slowest Liouvillian mode necessarily dominates the dynamics, the optimal interrogation time is governed by the Liouvillian gap and consequently exhibits the familiar inverse-gap scaling. 
At transient times, the coefficients $A^{\mu\nu}_{mnkl}(t)$ evolve together with the density matrix, allowing parameter-sensitive modes other than the slowest decaying one to contribute significantly to the QFIM. As a result, the transient optimum need not coincide with the asymptotic optimum, even in situations where an approximate inverse-gap scaling is observed. We shall return to this point later

 Retaining two dominant modes gives
\begin{equation}
F_{\mu\nu}(t)
\simeq
A_1(t)
t^2
e^{-2\Delta_1 t}
+
A_2(t)
t^2
e^{-2\Delta_2 t}.
\label{eq:twomode}
\end{equation}
The parameter indices of the Liouvillian mode coupling tensors have been suppressed for clarity. The stationarity condition becomes
\begin{align}
&
\dot A_1(t)
t^2
e^{-2\Delta_1t}
+
A_1(t)
\frac{d}{dt}
\left(
t^2
e^{-2\Delta_1t}
\right)
\nonumber\\
&+
\dot A_2(t)
t^2
e^{-2\Delta_2t}
+
A_2(t)
\frac{d}{dt}
\left(
t^2
e^{-2\Delta_2t}
\right)
=0.
\label{eq:twomodestationary}
\end{align}

Equation~(\ref{eq:twomodestationary}) shows that the optimal interrogation time depends jointly on the evolution of the modal coupling coefficients and the Liouvillian decay rates. Only when the Liouvillian gap overwhelmingly dominates and the coefficients vary slowly compared with the exponential kernels does Eq.~(\ref{eq:twomodestationary}) reduce to the inverse-gap result. Otherwise, the optimal interrogation time becomes a nonlinear function of the complete Liouvillian spectrum together with the time-dependent modal couplings, naturally leading to deviations from the conventional steadystate scaling relation $t_{\mathrm{opt}}\propto\Delta^{-1}$.

Writing the Liouvillian eigenvalues as $
\lambda_n=-\gamma_n+i\omega_n,
\gamma_n\ge\Delta
$, every exponential factor can be written as
\begin{equation}
e^{\lambda_nt}
=
e^{-\Delta t}
e^{-(\gamma_n-\Delta)t}
e^{i\omega_nt}.
\end{equation}
Consequently, every temporal kernel possesses the exact factorization
\begin{equation}
f_{mn}(t)
=
e^{-\Delta t}
\tilde f_{mn}(t),
\label{eq:kernel_factor}
\end{equation}
where
\begin{equation}
\tilde f_{mn}(t)
=
\begin{cases}
\dfrac{
e^{-(\gamma_m-\Delta)t+i\omega_mt}
-
e^{-(\gamma_n-\Delta)t+i\omega_nt}
}
{\lambda_m-\lambda_n},
&
m\neq n,
\\[2ex]
t\,e^{-(\gamma_n-\Delta)t+i\omega_nt},
&
m=n.
\end{cases}
\end{equation}
Substituting Eq.~(\ref{eq:kernel_factor}) into Eq.~(\ref{eq:qfim_modal}), THE QFIM takes a simpler form
\begin{equation}
F_{\mu\nu}(t)
=
e^{-2\Delta t}
G_{\mu\nu}(t),
\label{eq:qfi_factorized}
\end{equation}
where
\begin{equation}
G_{\mu\nu}(t)
=
\sum_{m,n}
\sum_{k,l}
A^{\mu\nu}_{mnkl}(t)
\tilde f_{mn}(t)
\tilde f_{kl}(t).
\label{eq:G_def}
\end{equation}
Equation~(\ref{eq:qfi_factorized}) can be understood as a separator of the universal exponential envelope determined by the Liouvillian gap from the remaining transient dynamics. The latter are contained entirely in the collective function $G_{\mu\nu}(t)$, which incorporates both the multimode spectral interference through the reduced kernels and the evolving statistical metric through the time-dependent coefficients $A^{\mu\nu}_{mnkl}(t)$.  Unlike the exponential envelope, $G_{\mu\nu}(t)$ is a genuinely multimode quantity. It incorporates the reduced temporal kernels $\tilde f_{mn}(t)$, which describe the relative evolution and interference of all Liouvillian relaxation channels after the common exponential decay has been removed, together with the time-dependent coefficients $A^{\mu\nu}_{mnkl}(t)$, which arise from the instantaneous quantum statistical metric. Consequently, $G_{\mu\nu}(t)$ characterizes how the surviving parameter information is distributed among the Liouvillian eigenspaces and how efficiently each mode contributes to parameter estimation at a given instant of time. This interpretation naturally separates the roles played by the Liouvillian spectrum. The Liouvillian gap determines the global rate at which information is lost from the system, whereas the remaining spectral modes govern the internal redistribution of the information that survives. The latter process includes both the interference between different relaxation channels and the continuous evolution of the statistical distinguishability of nearby quantum states as the density operator relaxes toward the stationary state. Thus, even after extracting the common exponential decay, the QFIM remains a genuinely multimode quantity throughout the transient evolution. Differentiating Eq.~(\ref{eq:qfi_factorized}) gives

\begin{equation}
\frac{dF_{\mu\nu}}{dt}
=
e^{-2\Delta t}
\left[
\dot G_{\mu\nu}(t)
-
2\Delta
G_{\mu\nu}(t)
\right].
\label{eq:stationary_env}
\end{equation}
Since the exponential prefactor is non-vanishing for finite times, the optimal interrogation time is determined by
\begin{equation}
\frac{\dot G_{\mu\nu}(t)}
{G_{\mu\nu}(t)}
=
2\Delta.
\label{eq:G_condition}
\end{equation}
Equation~(\ref{eq:G_condition}) provides a transparent interpretation of the transient metrological dynamics. The Liouvillian gap fixes the universal decay rate of the QFIM through the exponential envelope, whereas the quantity
$
\dot G_{\mu\nu}(t)/
G_{\mu\nu}(t)
$ measures the relative temporal evolution of the collective multimode contribution. Importantly, this collective contribution contains not only the interference among different Liouvillian relaxation modes but also the evolution of the quantum statistical metric as the density operator relaxes towards the stationary state. The maximum of the QFIM is therefore reached when the transient redistribution of parameter information among the Liouvillian modes balances its irreversible loss to the environment. Consequently, appearance of the an apparent inverse-gap scaling does not require the QFIM to be dominated by a single Liouvillian mode. Several relaxation modes may contribute simultaneously to the transient dynamics provided that their combined evolution remains slow compared with the universal decay rate set by the Liouvillian gap. Conversely, deviations from $t_{\mathrm{opt}}\propto\Delta^{-1}$ arise when the collective transient evolution encoded in $G_{\mu\nu}(t)$ becomes comparable to the universal exponential decay. Optimal interrogation time is
determined not only by the lifetime of the metrological
information but also by the rate at which that information is redistributed among the Liouvillian eigenspaces. The departure from inverse-gap scaling therefore reflects the increasing influence of the full transient Liouvillian dynamics, rather than simply the presence of multiple relaxation modes.

For large Liouvillians, in practice, the steady-state interrogation time is obtained numerically as the first time at which the QFI becomes indistinguishable from its steady-state value within a prescribed numerical tolerance. Denoting the steady-state QFIM by
\begin{equation}
F_{\mu\nu}^{(\infty)}
=
\lim_{t\rightarrow\infty}
F_{\mu\nu}(t),
\end{equation}
the numerical steady-state time $t_\infty$ is defined implicitly through
\begin{equation}
\left|
F_{\mu\nu}(t_\infty)
-
F_{\mu\nu}^{(\infty)}
\right|
=
\varepsilon ,
\label{eq:threshold}
\end{equation}
where $\varepsilon$ is the chosen convergence threshold.
Using the exact modal decomposition, $
F_{\mu\nu}(t)
=
\sum_{m,n}
\sum_{k,l}
A^{\mu\nu}_{mnkl}(t)
f_{mn}(t)
f_{kl}(t),
$
the deviation from the steady state becomes
\begin{align}
\Delta F_{\mu\nu}(t)
&=
F_{\mu\nu}(t)
-
F_{\mu\nu}^{(\infty)}
\nonumber\\
&=
\sum_{m,n}
\sum_{k,l}
\Big[
A^{\mu\nu}_{mnkl}(t)
f_{mn}(t)
f_{kl}(t)
-
A^{\mu\nu(\infty)}_{mnkl}
f^{(\infty)}_{mn}
f^{(\infty)}_{kl}
\Big].
\label{eq:deltaF}
\end{align}
Since both the temporal kernels and the statistical metric evolve during the relaxation,
the coefficients may be written as
\begin{equation}
A^{\mu\nu}_{mnkl}(t)
=
A^{\mu\nu(\infty)}_{mnkl}
+
\delta A^{\mu\nu}_{mnkl}(t),
\end{equation}
which gives
\begin{align}
\Delta F_{\mu\nu}(t)
&=
\sum_{m,n}
\sum_{k,l}
A^{\mu\nu(\infty)}_{mnkl}
\,
\Delta\!\left[
f_{mn}(t)
f_{kl}(t)
\right]
\nonumber\\
&\qquad
+
\sum_{m,n}
\sum_{k,l}
\delta A^{\mu\nu}_{mnkl}(t)
f_{mn}(t)
f_{kl}(t),
\label{eq:deltaF2}
\end{align}
where
\begin{equation}
\Delta\!\left[
f_{mn}(t)
f_{kl}(t)
\right]
=
f_{mn}(t)f_{kl}(t)
-
f^{(\infty)}_{mn}
f^{(\infty)}_{kl}.
\end{equation}

Equation~(\ref{eq:deltaF2}) shows that the approach of the QFIM towards its steady-state value is governed by two distinct relaxation mechanisms. The first originates from the decay of the Liouvillian temporal kernels, while the second arises from the time dependence of the quantum statistical metric through the coefficients
$A^{\mu\nu}_{mnkl}(t)$. Consequently, the convergence of the QFI is determined by the collective evolution of all parameter-sensitive Liouvillian eigenspaces rather than by the Liouvillian gap alone.

Substituting Eq.~(\ref{eq:deltaF2}) into the threshold condition
(\ref{eq:threshold}) shows that the numerically determined steady-state time satisfies
\begin{equation}
\left|
\sum_{m,n}
\sum_{k,l}
A^{\mu\nu(\infty)}_{mnkl}
\Delta(f_{mn}f_{kl})
+
\sum_{m,n}
\sum_{k,l}
\delta A^{\mu\nu}_{mnkl}
f_{mn}f_{kl}
\right|
=
\varepsilon .
\label{eq:threshold2}
\end{equation}
Unlike the asymptotic decay of an individual Liouvillian mode, Eq.~(\ref{eq:threshold2}) depends on the complete multimode relaxation of both the temporal kernels and the statistical metric. Therefore, the numerically extracted steady-state interrogation time need not exhibit inverse-gap scaling, even though the asymptotic relaxation of the density operator itself is ultimately governed by the Liouvillian gap.
%%%%%
%%%%%%

The exact decomposition of the QFIM contains contributions from all parameter-sensitive Liouvillian eigenspaces. To illustrate how the interplay between modal relaxation and statistical weighting determines the optimal interrogation time, it is sufficient to consider the lowest-order nontrivial truncation in which only two effective Liouvillian modes contribute appreciably. Restricting Eq.~(\ref{eq:qfim_modal}) to these modes gives
\begin{equation}
F(t)
=
A_{11}(t)f_1^2(t)
+
A_{22}(t)f_2^2(t)
+
2A_{12}(t)f_1(t)f_2(t),
\label{eq:twomode_general}
\end{equation}
where, for simplicity, the parameter indices have been suppressed and
$
f_i(t)
=
t\,e^{-\gamma_i t},
\qquad
\gamma_1=\Delta<\gamma_2,
$
correspond to the diagonal Liouvillian kernels.
Equation~(\ref{eq:qfim_modal}) therefore becomes
\begin{align}
F(t)
=
&
A_{11}(t)t^2e^{-2\Delta t}
+
A_{22}(t)t^2e^{-2\gamma_2t}
\nonumber\\
&
+
2A_{12}(t)t^2
e^{-(\Delta+\gamma_2)t}.
\label{eq:twomode_qfi}
\end{align}
The coefficients $A_{ij}(t)$ contain the instantaneous quantum statistical metric through the time-dependent density matrix and therefore generally evolve during the transient dynamics. Their role is to determine the statistical importance of each Liouvillian relaxation channel, whereas the exponential kernels determine the corresponding relaxation rates.
For transient dynamics, the optimal interrogation time satisfies
$
{\dot F(t)}=0.
\label{eq:toy_stationary}
$If the statistical metric varies slowly over the transient interval, the coefficients may be regarded as approximately constant over the optimization window. Equation~(\ref{eq:toy_stationary}) then becomes
\begin{align}
0
=
&
A_{11}
e^{-2\Delta t}
(1-\Delta t)
\nonumber\\
&
+
A_{22}
e^{-2\gamma_2t}
(1-\gamma_2t)
\nonumber\\
&
+
A_{12}
e^{-(\Delta+\gamma_2)t}
\left(
2-(\Delta+\gamma_2)t
\right).
\label{eq:toy_condition}
\end{align}

Several limiting cases immediately follow.
If
\begin{equation}
A_{11}
\gg
A_{22},A_{12},
\end{equation}
the slowest relaxation mode dominates the parameter encoding, and Eq.~(\ref{eq:toy_condition}) reduces to
\begin{equation}
1-\Delta t=0,
\end{equation}
yielding the conventional inverse-gap scaling
\begin{equation}
t^*
=
\frac{1}{\Delta}.
\end{equation}

When the statistical weights become comparable,
\begin{equation}
A_{11}
\sim
A_{22},
\end{equation}
the faster relaxation mode contributes on an equal footing with the gap mode. The stationarity condition then depends simultaneously on the statistical weights and both relaxation rates, so that the optimal interrogation time is no longer determined solely by the Liouvillian gap,
\begin{equation}
t^*
=
f(A_{11},A_{22},A_{12},\Delta,\gamma_2),
\end{equation}
leading naturally to deviations from inverse-gap scaling.

Conversely, if the relaxation rates satisfy
\begin{equation}
\gamma_2
\simeq
\Delta,
\end{equation}
the exponential kernels evolve on nearly identical timescales. Although several Liouvillian modes contribute with comparable statistical weight, Eq.~(\ref{eq:toy_condition}) remains approximately controlled by the common exponential envelope. Consequently,
\begin{equation}
t^*
\propto
\frac{1}{\Delta},
\end{equation}
may still hold despite the absence of single-mode dominance. This illustrates that an approximately linear dependence of the optimal interrogation time on the inverse Liouvillian gap should not be interpreted as evidence for unimodal dynamics.

The same decomposition also explains the behaviour of the numerically determined steady-state interrogation time. Defining the steady state through a numerical convergence threshold,
\begin{equation}
|F(t_\infty)-F_\infty|
=
\varepsilon,
\label{eq:toy_threshold}
\end{equation}
where
\begin{equation}
F_\infty
=
\lim_{t\rightarrow\infty}F(t),
\end{equation}
Eq.~(\ref{eq:twomode_qfi}) yields
\begin{align}
\varepsilon
=
&
A_{11}^{(\infty)}
e^{-2\Delta t_\infty}
+
A_{22}^{(\infty)}
e^{-2\gamma_2t_\infty}
\nonumber\\
&
+
2A_{12}^{(\infty)}
e^{-(\Delta+\gamma_2)t_\infty},
\label{eq:toy_threshold2}
\end{align}
where the polynomial prefactor has been omitted since the exponential relaxation dominates the long-time behaviour. Unlike the transient optimization condition, Eq.~(\ref{eq:toy_threshold2}) is a transcendental equation involving several relaxation channels simultaneously. Consequently, the threshold-defined steady-state interrogation time depends on the complete multimode relaxation process,
\begin{equation}
t_\infty
=
g(A_{11}^{(\infty)},
A_{22}^{(\infty)},
A_{12}^{(\infty)},
\Delta,
\gamma_2,
\varepsilon),
\end{equation}
and therefore need not scale linearly with the inverse Liouvillian gap, even though the asymptotic relaxation of the density operator is ultimately governed by the slowest Liouvillian eigenvalue.

Therefore, based on EQ. (38), Eq. (41), Eq. (56) and Eq. (62), the  appearance of linearity or  nonlinearity between the Liouvillian gap and optimal sensing time is definitely not indicative of the type of participation from the different spectral modes of the Liouvillian. The relationship can take any form and is not dependent on the number of modes participating.

%%%%%%%%%%
\section{Conclusion}
We  investigated the dynamical estimation of bath-induced coherence in an open quantum V-type system by  evaluating the transient quantum Fisher information matrix.  Our results demonstrate that the highest estimation precision is achieved during the transient evolution, whereas at the steady state, the estimation of coherence becomes metrologically inactive even though finite coherence persists. This establishes that the presence of coherence in a system out of equilibrium does not guarantee its measurement sensitivity. The obtained higher values of the quantum Fisher information indicate that coherences due to the hot (cold) bath have better precision at higher (lower) temperatures.  

We then analyze the transient optimization of the quantum Fisher information through the lens of the Liouvillian spectrum.
A central finding of this work is that the optimal estimation time is not, in general, determined by the inverse Liouvillian gap. Although the relaxation gap accurately characterizes the asymptotic approach to the steady state, it does not uniquely determine the timescale over which parameter information is most effectively encoded. Instead, the metrological dynamics arise from the combined evolution of several Liouvillian eigenmodes. We  identified which modes actually contribute to the transient peak not on the basis of the magnetude but on the basis of the full time-dependent density matrix's modal expansion amplitudes. These statistical weights are not too sensitive to coeherences. 

There is a complete absence of universal scaling behavior of the optimal interrogation time with the inverse Liouvillian gap. Participation from multiple eigenmodes can lead to linear scaling and yet unimodal participation can lead to nonlinear scaling. This is due to the competition of faster modes with the slowest relaxation mode during transient dynamics through the spectrally decomposed parameters, namely the relaxation eigenvalues and the modal statistical weights.  Even though the slowest relaxation timescale is determined by the Liouvillian gap, faster relaxation channel are capable of providing dominant contribution to the emergence of the transient optimum at lower temperatures. Neither is the emergence of nonlinear scaling, by itself, an indicator of genuine multimodal participation of Liouvillian spectrum to the transient optimization, not is linear behavior an indicator of slowest mode participation during transient times. 

Similar qualitative behavior is observed in the off-diagonal element of the quantum Fisher information matrix. Therefore, the underlying principle is robust in both single- and multi-parameter estimation. By considering  a general Markovian equation, we analytically proved why the optimal time can be linear or non-linear with respect to Liouvillian gap and it is not indicative of any multimodal or unimodal participation. We achieved this by spectrally decomposing the time-dependent quantum Fisher information matrix into temporal kernels determined by the Liouvillian spectrum (a spectral contribution) and modal coupling tensors (statistical contribution) that encode the parameter sensitivity of the dynamics together with the instantaneous quantum statistical metric. A simple analysis reveal when and how linearity and nonlinearity follows without the appearance of any universal scaling behavior with respect to the Liouvillian gap.

Finally, we showed that coupling the finite system to a quantum cavity coupling provides an effective means of controlling the estimation dynamics, albeit from a perturbative approach. When the system-cavity coupling remains weaker than the system-bath interaction, the metrological behavior is essentially unchanged and remains transient. However, when coherent coupling dominates and is assisted by a pronounced thermal bias between the hot and cold baths, the transient metrological enhancement shifts to the steadystate, where the Fisher information for coherence is larger.  Our results establish a clear distinction between relaxation timescales and metrological timescales during coherence estimation in open quantum systems. Coherence estimation estimation cannot, in general, be understood solely in terms of the Liouvillian gap. 

%%%%%%%%%%%%%%%%%%%%%%%%%%%%%%%%%%%%%%%%%%%%%%%%%%%%%%%%%%%%%

%%%%%%%%%%%%%%%%%%%%%%%%%%

%%%%%%%%%%%%%
\bibliographystyle{unsrt} 
\bibliography{Reference}

@article{Alipour2014,
  author = {Alipour, S. and Mehboudi, M. and Rezakhani, A. T.},
  title = {Quantum Metrology in Open Systems: Dissipative Cram\'er-Rao Bound},
  journal = {Physical Review Letters},
  volume  = {112},
  number = {12},
  pages = {120405},
  year = {2014},
  doi = {10.1103/PhysRevLett.112.120405},
  
}

@article{Braunstein1994,
  author = {Braunstein, Samuel L. and Caves, Carlton M.},
  title = {Statistical Distance and the Geometry of Quantum States},
  journal = {Physical Review Letters},
  volume = {72},
  number = {22},
  pages = {3439--3443},
  year = {1994},
  doi = {10.1103/PhysRevLett.72.3439},
  
}

@article{Camati2019,
  author = {Camati, Patrice A. and Santos, Jonas F. G. and Serra, Roberto M.},
  title = {Coherence Effects in the Performance of the Quantum {{Otto}} Heat Engine},
  journal = {Physical Review A},
  volume = {99},
  number = {6},
  pages = {062103},
  year = {2019},
  doi = {10.1103/PhysRevA.99.062103},
  
}

@article{Diehl2008,
  author = {Diehl, S. and Micheli, A. and Kantian, A. and Kraus, B. and B{\"u}chler, H. P. and Zoller, P.},
  title = {Quantum States and Phases in Driven Open Quantum Systems with Cold Atoms},
  journal = {Nature Physics},
  volume = {4},
  number = {11},
  pages = {878--883},
  year = {2008},
  doi = {10.1038/nphys1073},
  
}

@article{Dorfman2018,
  author = {Dorfman, Konstantin E. and Xu, Dazhi and Cao, Jianshu},
  title = {Efficiency at Maximum Power of a Laser Quantum Heat Engine Enhanced by Noise-Induced Coherence},
  journal = {Physical Review E},
  volume = {97},
  number = {4},
  pages = {042120},
  year = {2018},
  doi = {10.1103/PhysRevE.97.042120},
  
}

@article{Dorfman2013,
  author = {Dorfman, Konstantin E. and Voronine, Dmitri V. and Mukamel, Shaul and Scully, Marlan O.},
  title = {Photosynthetic Reaction Center as a Quantum Heat Engine},
  journal = {Proceedings of the National Academy of Sciences},
  volume = {110},
  number = {8},
  pages = {2746--2751},
   year = {2013},
  doi = {10.1073/pnas.1212666110},
 
}

@article{Escher2011,
  author = {Escher, B. M. and De Matos Filho, R. L. and Davidovich, L.},
  title = {General Framework for Estimating the Ultimate Precision Limit in Noisy Quantum-Enhanced Metrology},
  journal = {Nature Physics},
  volume = {7},
  number = {5},
  pages = {406--411},
  year = {2011},
  doi = {10.1038/nphys1958},
}

@article{Giovannetti2011,
  author = {Giovannetti, Vittorio and Lloyd, Seth and Maccone, 
  Lorenzo},
  title = {Advances in Quantum Metrology},
  journal = {Nature Photonics},
  volume = {5},
  number = {4},
  pages = {222--229},
  year = {2011},
  doi = {10.1038/nphoton.2011.35},
}

@article{Giovannetti2004,
  author = {Giovannetti, Vittorio and Lloyd, Seth and Maccone, Lorenzo},
  title = {Quantum-Enhanced Measurements: Beating the Standard Quantum Limit},
  journal = {Science},
  volume = {306},
  number = {5700},
  pages = {1330--1336},
  year = {2004},
  doi = {10.1126/science.1104149},
}

@article{Gu2010,
  author = {Gu, Shi-Jian},
  title = {FIDELITY APPROACH TO QUANTUM PHASE TRANSITIONS},
  journal = {International Journal of Modern Physics B},
  volume = {24},
  number = {23},
  pages = {4371--4458},
  year = {2010},
  doi = {10.1142/S0217979210056335},
}

@article{Kessler2012,
  author = {Kessler, E. M. and Giedke, G. and Imamoglu, A. and Yelin, S. F. and Lukin, M. D. and Cirac, J. I.},
  title = {Dissipative Phase Transition in a Central Spin System},
  journal = {Physical Review A},
  volume = {86},
  number = {1},
  pages = {012116},
   year = {2012},
  doi = {10.1103/PhysRevA.86.012116},
}

@article{Minganti2018,
  author = {Minganti, Fabrizio and Biella, Alberto and Bartolo, Nicola and Ciuti, Cristiano},
  title = {Spectral Theory of Liouvillians for Dissipative Phase Transitions},
  journal = {Physical Review A},
  volume = {98},
  number = {4},
  pages = {042118},
  year = {2018},
  doi = {10.1103/PhysRevA.98.042118},
}

@article{Mori2020,
  author = {Mori, Takashi and Shirai, Tatsuhiko},
  title = {Resolving a Discrepancy between Liouvillian Gap and Relaxation Time in Boundary-Dissipated Quantum Many-Body Systems},
  journal = {Physical Review Letters},
  volume = {125},
  number = {23},
  pages = {230604},
  year = {2020},
  doi = {10.1103/PhysRevLett.125.230604},
}

@article{Mori2023,
  author = {Mori, Takashi and Shirai, Tatsuhiko},
  title = {Symmetrized Liouvillian Gap in Markovian Open Quantum Systems},
  journal = {Physical Review Letters},
  volume = {130},
  number = {23},
  pages = {230404},
   year = {2023},
  doi = {10.1103/PhysRevLett.130.230404},
}

@article{Paris2009,
  author = {Paris, Matteo G. A.},
  title = {QUANTUM ESTIMATION FOR QUANTUM TECHNOLOGY},
  journal = {International Journal of Quantum Information},
  volume = {07},
  number = {supp01},
  pages = {125--137},
  year = {2009},
  doi = {10.1142/S0219749909004839},
}

@article{Peng2024,
  author = {Peng, Jia-Xin and Zhu, Baiqiang and Zhang, Weiping and Zhang, Keye},
  title = {Dissipative Quantum Fisher Information for a General Liouvillian Parametrized Process},
  journal = {Physical Review A},
  volume = {109},
  number = {1},
  pages = {012432},
   year = {2024},
  doi = {10.1103/PhysRevA.109.012432},
}

@article{Pezze2018,
  author = {Pezz{\`e}, Luca and Smerzi, Augusto and Oberthaler, Markus K. and Schmied, Roman and Treutlein, Philipp},
  title = {Quantum Metrology with Nonclassical States of Atomic Ensembles},
  journal = {Reviews of Modern Physics}, 
  volume = {90},
  number = {3},
  pages = {035005},
   year = {2018}, 
  doi = {10.1103/RevModPhys.90.035005},
}

@article{Poyatos1996,
  author = {Poyatos, J. F. and Cirac, J. I. and Zoller, P.},
  title = {Quantum Reservoir Engineering with Laser Cooled Trapped Ions},
  journal = {Physical Review Letters},
  volume = {77},
  number = {23},
  pages = {4728--4731},
  year = {1996},
  doi = {10.1103/PhysRevLett.77.4728},
}

@article{Sarmah2024,
  author = {Sarmah, Manash Jyoti and Goswami, Himangshu Prabal},
  title = {Noise-Induced Coherent Ergotropy of a Quantum Heat Engine},
  journal = {Physical Review A},
  volume = {110},
  number = {3},
  pages = {032213},
  year = {2024},
  doi = {10.1103/PhysRevA.110.032213},
}

@article{Scully2003,
  author = {Scully, Marlan O. and Zubairy, M. Suhail and Agarwal, Girish S. and Walther, Herbert},
  title = {Extracting Work from a Single Heat Bath via Vanishing Quantum Coherence},
  journal = {Science},
  volume = {299},
  number = {5608},
  pages = {862--864},
  year = {2003},
  doi = {10.1126/science.1078955},
}

@article{Scully2011,
  author = {Scully, Marlan O. and Chapin, Kimberly R. and Dorfman, Konstantin E. and Kim, Moochan Barnabas and Svidzinsky, Anatoly},
  title = {Quantum Heat Engine Power Can Be Increased by Noise-Induced Coherence},
  journal = {Proceedings of the National Academy of Sciences},
  volume = {108},
  number = {37},
  pages = {15097--15100},
   year = {2011},
  doi = {10.1073/pnas.1110234108}, 
}

@article{Uzdin2015,
  author = {Uzdin, Raam and Levy, Amikam and Kosloff, Ronnie},
  title = {Equivalence of Quantum Heat Machines, and Quantum-Thermodynamic Signatures},
  journal = {Physical Review X},
  volume = {5},
  number = {3},
  pages = {031044},
  year = {2015},
  doi = {10.1103/PhysRevX.5.031044},
}

@article{Verstraete2009,
  author = {Verstraete, Frank and Wolf, Michael M. and Ignacio Cirac, J.},
  title = {Quantum Computation and Quantum-State Engineering Driven by Dissipation},
  journal = {Nature Physics},
  volume = {5},
  number = {9},
  pages = {633--636},
   year = {2009},
  doi = {10.1038/nphys1342},
}

@article{Wang2014,
  author = {Wang, Teng-Long and Wu, Ling-Na and Yang, Wen and Jin, Guang-Ri and Lambert, Neill and Nori, Franco},
  title = {Quantum Fisher Information as a Signature of the Superradiant Quantum Phase Transition},
  journal = {New Journal of Physics},
  volume = {16},
  number = {6},
  pages = {063039},
   year = {2014},
  doi = {10.1088/1367-2630/16/6/063039},
}

@article{Yu2022,
  author = {Yu, Meng and Yang, Yang and Xiong, Hengna and Lin, Xianqin},
  title = {Critical Behavior of Quantum Fisher Information in Finite-Size Open Dicke Model},
  journal = {AIP Advances},
  volume = {12},
  number = {5},
  pages = {055118},
  year = {2022},
  doi = {10.1063/5.0091100},
}

@article{Demkowicz2015,
  author  = {Demkowicz-Dobrza{\'n}ski, Rafa{\l} and Jarzyna, Marcin and Ko{\l}ody{\'n}ski, Jan},
  title   = {Quantum Limits in Optical Interferometry},
  journal = {Progress in Optics},
  volume  = {60},
  pages   = {345--435},
  year    = {2015},
  doi     = {10.1016/bs.po.2015.02.003},
}

@misc{Midha2025,
  author = {Midha, Siddhant and Gopalakrishnan, Sarang},
  title  = {Metrology of Open Quantum Systems from Emitted Radiation},
  year   = {2025},
  note   = {arXiv:2504.13815 [quant-ph], \url{https://doi.org/10.48550/arXiv.2504.13815}},
}

@article{Vicentini2018,
  author = {Vicentini, Filippo and Minganti, Fabrizio and Rota, Riccardo and Orso, Giuliano and Ciuti, Cristiano},
  title = {Critical Slowing down in Driven-Dissipative Bose-Hubbard Lattices},
  journal = {Physical Review A},
  volume = {97},
  number = {1},
  pages = {013853},
  year = {2018},
  doi = {10.1103/PhysRevA.97.013853},
}

@article{Goswami2013,
  author = {Goswami, Himangshu Prabal and Harbola, Upendra},
  title = {Thermodynamics of quantum heat engines},
  journal = {Physical Review A},
  volume = {88},
  number = {1},
  pages = {013842},
  year = {2013},
  doi = {10.1103/PhysRevA.88.013842}
}

@article{Carollo2020,
  author = {Carollo, Angelo and Valenti, Davide and Spagnolo, Bernardo},
  title = {Geometry of Quantum Phase Transitions},
  journal = {Physics Reports},
  volume = {838},
  pages = {1--72},
  year = {2020},
  doi ={10.1016/j.physrep.2019.11.002},
}

@article{Thakuria2026,
  author = {Thakuria, Bitap Raj and Kalita, Trishna and Akhtar, Javed and Goswami, Himangshu Prabal},
  title = {Fisher Information of a Nonequilibrium Anharmonic Donor-Acceptor Rectifier},
  journal = {Physica A},
  volume = {697},
  pages = {131723},
   year = {2026},
  doi = {10.1016/j.physa.2026.131723},
}

@article{Zhou2022,
  author = {Zhou, Bozhen and Wang, Xueliang and Chen, Shu},
  title = {Exponential Size Scaling of the Liouvillian Gap in Boundary-Dissipated Systems with Anderson Localization},
  journal = {Physical Review B},
  volume = {106},
  number={6},
  pages = {064203},
  year = {2022},
  doi = {10.1103/PhysRevB.106.064203}
}

@article{Haga2021,
  author = {Haga, Taiki and Nakagawa, Masaya and Hamazaki, Ryusuke and Ueda, Masahito},
  title = {Liouvillian Skin Effect: Slowing Down of Relaxation Processes without Gap Closing},
  journal = {Physical Review Letters},
  volume = {127},
  number={7},
  pages = {070402},
  year = {2021},
  doi = {10.1103/PhysRevLett.127.070402}
}

@article{Creatore2013,
  author={Creatore, Celestino and Parker, M Andy and Emmott, Stephen and Chin, Alex W},
  title = {Efficient biologically inspired photocell enhanced by delocalized quantum states},
  journal = {Physical review letters},
  volume = {111},
  number = {25},
  pages = {253601},
  year = {2013},
   doi = {10.1103/PhysRevLett.111.253601},
}

@article{Saleem2023,
  author={Saleem, Zain H and Shaji, Anil and Gray, Stephen K},
  title={Optimal time for sensing in open quantum systems},
  journal = {Physical Review A},
  volume = {108},
  number = {2},
  pages = {022413},
  year = {2023}, 
  doi =  {10.1103/PhysRevA.108.022413},
}

@article{Holubec2018,
  author={Holubec, Viktor and Novotn{\`y}, Tom{\'a}{\v{s}}},
  title={Effects of noise-induced coherence on the performance of quantum absorption refrigerators},
  journal={Journal of Low Temperature Physics},
  volume={192},
  number={3},
  pages={147--168},
  year={2018},
  doi = {10.1007/s10909-018-1960-x},
 
}

@article{Holubec2019,
  author={Holubec, Viktor and Novotn{\`y}, Tom{\'a}{\v{s}}},
  title={Effects of noise-induced coherence on the fluctuations of current in quantum absorption refrigerators},
  journal={The Journal of chemical physics},
  volume={151},
  number={4},
  year = {2019},
  doi = {10.1063/1.5096275},
}

@article{sarmah2024efficiency,
  title={Efficiency fluctuations of a heat engine with noise-induced quantum coherences},
  author={Sarmah, Manash Jyoti and Goswami, Himangshu Prabal},
  journal={Physical Review A},
  volume={110},
  number={5},
  pages={052214},
  year={2024},
  doi = {10.1103/PhysRevA.110.052214},
  publisher={APS}
}

\end{document}